\documentclass[amsmath,amssymb,twocolumn,superscriptaddress]{revtex4}
\usepackage{graphicx}% Include figure files
\usepackage{dcolumn}% Align table columns on decimal point

\def\be{\begin{equation}}
\def\ee{\end{equation}}
\def\beq{\begin{eqnarray}}
\def\eeq{\end{eqnarray}}

\usepackage{bm}% bold math
\usepackage{graphicx}
\usepackage{multirow}
\usepackage{dcolumn}
\usepackage{amsmath}
\usepackage[latin1]{inputenc}
\usepackage{graphicx, psfrag}
\usepackage{amssymb}
\usepackage[colorlinks=true, citecolor=blue, urlcolor = blue, linkcolor= red, bookmarks=true]{hyperref}
\usepackage{float}
\usepackage{amsmath}
\usepackage{amsfonts}
\usepackage{dcolumn}
\usepackage{hyperref}
\usepackage{subfigure}
\usepackage{pgfplots}
\usepackage{epstopdf}
\usepackage{booktabs}

\begin{document}

%\preprint{}

\title{Anisotropic Strange Star Model Beyond Standard Maximum Mass Limit by Gravitational Decoupling in $f(Q)$ Gravity} 



\author{S. K. Maurya}%
\email[Email:]{sunil@unizwa.edu.om}
\affiliation{Department of Mathematical and Physical Sciences,
College of Arts and Sciences, University of Nizwa, Nizwa, Sultanate of Oman}

\author{Ksh. Newton Singh} 
\email[Email:]{ntnphy@gmail.com} 
\affiliation{Department of Physics, National Defence Academy, Khadakwasla, Pune 411023, India}

\author{Santosh V. Lohakare}
\email[Email:]{lohakaresv@gmail.com}
\affiliation{Birla Institute of Technology and Science-Pilani, Hyderabad Campus, Hyderabad 500078, India}

\author{B. Mishra}
\email[Email:]{bivu@hyderabad.bits-pilani.ac.in}
\affiliation{Birla Institute of Technology and Science-Pilani, Hyderabad Campus, Hyderabad 500078, India, }

%\thanks{}

\date{\today}

\begin{abstract}
 The current theoretical development identified as the gravitational decoupling via Complete Geometric Deformation (CGD) method that has been introduced to explore the nonmetricity $Q$ effects in relativistic astrophysics. In the present work, we have investigated the gravitationally decoupled anisotropic solutions for the strange stars in the framework of $f(Q)$ gravity by utilizing the CGD technique. To do this, we started with Tolman metric ansatz along with the MIT Bag model equation of state related to the hadronic matter. The solutions of the governing equations of motions are obtained by using two approaches, namely the mimicking of the $\theta$ sector to the seed radial pressure and energy density of the fluid model. The obtained models describe the self-gravitating static, compact objects whose exterior solution can be given by the vacuum Schwarzschild Anti-de Sitter spacetime. In particular, we modeled five stellar candidates, viz., LMC X-4, PSR J1614-2230, PSR J0740+6620, GW190814, and GW 170817 by using the observational data. The rigorous viability tests of the solutions have been performed through regularity and stability conditions. We observed that the nonmetricity parameter and decoupling constant show a significant effect on stabilizing to ensure the physically realizable stellar models. The innovative feature of this work is to present the stable compact objects with the masses beyond the $2 M_{\odot}$ without engaging of exotic matter. Therefore, the present study shows a new perception and physical significance about the exploration of ultra-compact astrophysical objects.
\end{abstract}
%\keywords{}

\maketitle


\section{Introduction}  \label{sec1}
General theory of relativity (GR) has been formulated in the Riemannian space-time and from the physical and mathematical point of view, there are reasons to go beyond Riemannian space-time. Among many reasons, the present significant reason is the cosmic expansion and late time acceleration of the Universe \cite{Riess98,Perlmutter99}. Mathematically the non-Riemannian geometries, in particular the nonmetricity, can be obtained from the gauge theoretical \cite{Benn82,Hehl95} and group theoretical approaches \cite{Kirsch05,Boulanger06}. In the formulation of GR, Einstein attributes gravity to the space-time curvature and there are two other equivalent formulations that exist with the torsion and nonmetricity for the flat space-time. Another motivation on the nonmetricity approach is to improve the renormalizability of GR to reach to quantum gravitational theory.  So, the Einstein gravity can be described by the Einstein-Hilbert action as, $\int\sqrt{-g}\,R$, $\int\sqrt{-g}\,T$ and $\int\sqrt{-g}\,Q$ respectively for the curvature, torsion \cite{Aldrovandi13} and nonmetricity \cite{Jimenez18}. In the nonmetricity approach the curvature and torsion vanishes and the gravitational information is encoded from the contribution of nonmetricity only \cite{Nester99,Jimenez16,Golovnev17,Conroy18}. In addition, in this approach, the connection corresponds to the gauge potential and the metric represents the gravitational field \cite{Adak18}. Weyl \cite{Weyl18} attempted to make the unification of electromagnetism and gravitation with more general geometry in which the covariant divergence of the metric tensor is zero. Mathematically, this has been known as a new geometric quantity, known as nonmetricity. It was Einstein who criticized that, this theory is in contradiction with the known experimental results \cite{Einstein18}. Another extension to GR was proposed by Cartan who proposed the Einstein-Cartan theory, where the torsion field was proposed \cite{Cartan23}. Weitzenb$\ddot{o}$ck formulated an independent mathematical formulation, called as the Weitzenb$\ddot{\text{o}}$ck space \cite{Weitzenbock23}. The curvature of the Weitzenb$\ddot{\text{o}}$ck space is zero and the geometries have the property of teleparallelism.  As in geometry, the curvature has been replaced by the torsion, it is popularly known as TEGR (Teleparallel Equivalent of GR). So, GR can have curvature and torsion representation and in addition, another representation is nonmetricity, where the geometric variables are represented with the nonmetricity $Q$, known as the symmetric teleparallel gravity. To note in this representation, both the curvature and torsion remain zero. This has been further developed into $f(Q)$ gravity \cite{Jimenez18}. In this theory, under the teleparallelism constraint, one can always pick the coincident gauge, which prevents the affine connection from disappearing. TEGR and symmetric teleparallel gravity can be generalized to $f(T)$ gravity \cite{Cai16,Bahamonde21} and $f(Q)$ gravity \cite{Jimenez18,Heisenberg19}. Another important development is the $f(Q)$ cosmology was discussed by Jimenez \textit{et al} \cite{Jimenez20}. 

Some of the recent problems dealt in $f(Q)$ gravity have been discussed. The accelerated expansion of the Universe has been constructed without invoking exotic dark energy and the dynamical system has been investigated by Lu \textit{et al} \cite{Lu19}. The evidences on the deviation from $\Lambda$CDM model can be shown with the observational constraints \cite{Lazkoz19,Anagnostopoulos21}. Applications of spherically symmetric configuration has been studied in $f(Q)$ gravity \cite{Lin21}, while the perturbative corrections to the Schwarzschild solution have been discussed  by D'Ambrosio \cite{Ambrosio22}. The Schwarzschild-like solutions can also be obtained in nontrivial $f(Q,T)$ gravity, and its asymptotic behavior can also be studied Wang \textit{et al} \cite{Wang22}. It has been shown that nonmetricity can provide potentially observable effects in microscopic systems, which would help to impose tight constraints on the model parameters \cite{Latorre18}. The Noether symmetries approach has been explored in symmetric teleparallel gravity in order to reduce the dynamics of the system and hence to obtain the analytical solutions \cite{Dialektopoulos19}. To validate the nonmetricity gravity with the cosmological aspects, Barros \textit{et al} \cite{Barros20} have used the redshift space distortions data while performing the Bayesian statistical analysis. Constraining the functional form with the use of the order reduction method, the cosmological bounce scenario can be obtained from $f(Q)$ gravity field equations \cite{Bajardi20}. Another important remark on this gravity is that on large scale it delivers the dark energy aspects and on the galactic scale, it recovers Modified Newton Dynamics (MOND) \cite{Ambrosio20}. Some other aspects of the nonmetricity gravity works can be seen in Ref. and therein    \cite{Flathmann21,Ayuso21,Khyllep21,Fruscinate21,Capozziello:plb}. Another interesting aspect of this gravity is to obtain the model through a designer approach, that distinguishes it from the $\Lambda$CDM scenario \cite{Albuquerque22}.

To study the higher order corrections in self-gravitating compact objects, Einstein-Gauss-Bonnet (EGB) has been extensively used because it preserves the fundamental elements of General relativity such as the conservation laws via the Bianchi identities, diffeomorphism invariance, and quasi-linear, second order equations of motion. Even though the nonlinearity of the EGB field equations, several studies on the physical properties of compact objects such as surface redshift, stability, and the mass-to-radius ratio has been done in the context of 5D EGB formulation and compared their classical 5D counterparts \cite{hans1,hans2,hans3,hans4}. In this connection, Maharaj \textit{et al} recently \cite{hans6} proposed a new solution-generating algorithm via gravitational decoupling for isotropic matter distributions in the framework of 5 and 6-dimensional EGB gravity. On the other hand, another important development is the introduction of gravitational decoupling (GD) via the Minimal Geometric Deformation (MGD) approach in the EGB framework, from which the effects of anisotropic stress on small objects can be investigated. The first direct, simple, and systematic MGD approach was originally proposed by Ovalle \cite{OvallePRD2017} in the context of GR for modeling of compact objects, and later on it was extended in \cite{OvallePLB2019}. The modeling of compact objects using MGD and CGD has received a lot of attention in astrophysical research. Some of the recent research results in GR, $f(R,T)$ and $f(G)$ gravity theories can be seen in the references \cite{Panotopoulos18,Gabbanelli18,Contreras19,Ovalle:2018vmg,Gabbanelli:2019txr,Ovalle:2017wqi,daRocha:2021aww,daRocha:2021sqd,Estrada19,Maurya20,Zubair21a,Zubair21b,Muneer:2021lfz, Zubair:2021zqs, Azmat:2021qig,Contreras21,Leon21,MauryaFRT,SharifFG,Abellan:2020dze}. In this connection, a first systematic approach for MGD and extended MGD in 5D EGB gravity was proposed by Maurya \textit{et al} \cite{Maurya21a,Maurya21b} to obtain the exact solution of compact star model. Recently, Maurya \textit{et al} \cite{Maurya22} have shown the impact of the decoupling parameter and the EGB constant on the maximum mass limit on minimally deformed neutron star models. Furthermore, the gravitational decoupling approach has been successfully applied to discuss the complexity of the self-gravitating system \cite{Casadio:2019usg,Contreras:2022vmk,Carrasco-Hidalgo:2021dyg,Contreras:2021xkf,Maurya:2022cyv,Maurya:2022pef,Maurya:2021yhc}. 

Due to the success of this gravitational decoupling methodology in a different kind of gravity, we intended to apply this technique in $f(Q)$ gravity theory to model the astrophysical compact objects beyond the standard maximum mass limit. 

The $f(Q)$ theory of gravity has become a popular choice amongst cosmologists in explaining the late-time acceleration of the Universe without appealing to exotic matter distributions. In this work, our motivation is to study the complete deformed strange star model in symmetric teleparallel $f(Q)$ gravitational theory. The interest is to model the secondary compact object, which can be performed through the gravitational waves suggested by the LIGO-Virgo group. The investigation was made on the role of anisotropy in the observed radius and mass of the secondary component of event GW190814. Here, the data from GW190814 are used to determine the completely deformed strange star model in symmetric teleparallel $f(Q)$ gravity. To note here,  LIGO Hanford, LIGO Livingston, and LIGO Virgo were operating with average O3 sensitivity at GW 190814. Abbott \textit{et al} \cite{Abbott19, Abbott20a, Abbott20b} claim that this is the most substantial limit on the main spin to be smaller than 0.07. Further, it has been estimated that a merger rate density of $1-23 ~~ Gpc^{-3} yr^{-1}$ for GW 190814-like occurrences if we treat it as a novel class of compact binary coalescences. The rapid formation of compact binaries with this uncommon mass combination challenges our current understanding of astrophysical models.

The paper is organized as follows: In Section \ref{sec2}, the field equations of $f(Q)$ gravity with extra source has been derived. The gravitational decoupled solution in the nonmetricity gravity has been derived in Section \ref{sec3}. In Section \ref{sec4}, the boundary conditions are mentioned along with the graphical behaviors of the pressure, energy density, and other relevant parameters. The physical analysis is given in Section \ref{sec5} and finally, the conclusions are made on the problem investigated in Section \ref{sec6}. Some lengthy physical quantity statements have been consigned to the Appendix.

\section{Field equations for $f(Q)$ gravity with extra source}\label{sec2}
This paper will give a complete description of the symmetric teleparallel $f(Q)$ gravity for gravitationally decoupled systems. The gravitational interaction is generated by the nonmetricity scalar $Q$ in $f(Q)$ gravity. By incorporating an additional Lagrangian $\mathcal{L}_\theta$ for another source $\theta_{ij}$, the action for modified $f(Q)$ gravity for gravitationally decoupled systems can be stated as:
\begin{equation}\label{1}
\mathcal{S}=\int{\left(\frac{1}{2}\,f(Q)+\mathcal{L}_m+\beta \mathcal{L}_\theta\right)}\,\sqrt{-g}~d^4x\, ,
\end{equation}
where $g$ represents the determinant of the metric tensor $g_{ij}$, $\mathcal{L}_m$ describes the matter Lagrangian density, and $\beta$ is a decoupling constant. The $f(Q)$ gravity addresses a generic metric-affine spacetime, in which the metric tensor $g_{ij}$ and the connection $\Gamma^k_{\,\,\,i j}$ are considered separately, and the nonmetricity of the connection is defined by
\begin{equation}\label{2}
Q_{k i j}=\bigtriangledown_{_k}\,g_{i j}=\partial_i\, g_{jk}-\Gamma^l_{\,\,\,i j}\, g_{l k}-\Gamma^l_{\,\,\,i k} \,g_{j l},
\end{equation}
where $\nabla_{k}$ is the covariant derivative and $\Gamma^k_{\,\,\,i j}$ is known as affine connection, the following three independent components can be reduced into the generic form of affine connection:
\begin{equation}\label{3}
\Gamma^k_{i j}=\lbrace^{\,\,\,k}_{\,i\,\,j} \rbrace+K^k_{\,\,\,ij}+ L^k_{\,\,ij},
\end{equation}
where $\lbrace^{\,\,\,k}_{\,i\,\,j} \rbrace$, $L^k_{\,\,ij}$, and $K^k_{\,\,\,ij}$ are the Levi-Civita connection, deformation tensor and contortion tensor respectively, which are defined as follow:
\begin{eqnarray}
\label{4}
&&\hspace{-0.2cm} \lbrace^{\,\,\,k}_{\,i\,\,j} \rbrace=\frac{1}{2}g^{ kl}\left(\partial_i g_{l j}+\partial_ j g_{l i}-\partial_l g_{i j}\right),~~\nonumber\\&&\hspace{-0.2cm}
L^k_{\,\,\,i j}=\frac{1}{2}Q^k_{\,\,\,i j}-Q_{(i \,\,\,\,\,\,j)}^{\,\,\,\,\,\,k},~~~ K^k_{\,\,\,ij}=\frac{1}{2} T^k_{\,\,\,i j}+T_{(i \,\,\,\,\,\,j)}^{\,\,\,\,\,\,k},
\end{eqnarray}
with the anti-symmetric component of the affine connection defined by the torsion tensor $T^k_{\,\,\,i \,j}=2\Gamma^l_{\,\,\,[i\,j]}$. The superpotential related to the nonmetricity tensor is defined as:
\begin{eqnarray}
\label{5}
P^k_{\,\,\,\,ij}=\frac{1}{4}\left[-Q^k_{\,\,\,\,ij}+2 Q_{(i\,j)}^k+Q^k g_{ij}-\tilde{Q}^k g_{ij}-\delta^k_{(i}Q_{j)}\right],~~~
\end{eqnarray}
We can only have two independent traces from the nonmetricity tensor $Q_{k i j}$ because of the symmetry of the metric tensor $g_{i j}$, namely,
\begin{eqnarray}
Q_{k}\equiv Q_{k\,\,\,~i}^{\,\,~i},~~~~~~~~\; \tilde{Q}^k=Q^{\,\,\,\,ki}_{i}.
\end{eqnarray}
Let us now present the nonmetricity scalar, which will be important in our work,
\begin{equation}\label{7}
Q=-Q_{ijk}\,P^{ijk}=-g^{m n}\left(L^i_{\,\,\,jn} L^j_{\,\,\,m i}-L^j_{\,\,\,ij} L^i_{\,\,\,m n}\right).
\end{equation}
To compute the field equations for $f(Q)$ gravity, we may make the action Equation (\ref{1}) constant in terms of variation across the metric tensor $g_{i j}$, obtaining:
\begin{eqnarray}
\label{8}
\hspace{-0.2cm}\frac{2}{\sqrt{-g}}\bigtriangledown_k\left(\sqrt{-g}\,f_Q\,P^k_{\,\,\,\,i j}\right)+\frac{1}{2}g_{i j}f 
+f_Q\big(P_{i\,k l}\,Q_j^{\,\,\,\,k l}-2\,Q_{k l i}\,P^{k l}_{\,\,\,\,\,j}\big) \nonumber\\=- T^{\text{eff}}_{i j}, ~
~~~\text{where}~~T^{\text{eff}}_{i j}=\big(T_{i j}+c \,\theta_{i j}\big),~~~~
\end{eqnarray}
where a subscript $Q$ represents a derivative of $f(Q)$ with respect to $Q$,  $f_Q=\frac{\partial f}{\partial Q}$, and $T_{i j}$ is the energy-momentum tensor and extra source $\theta_{i j}$, whose forms are respectively,
\begin{eqnarray}
\label{9}
T_{ij}=-\frac{2}{\sqrt{-g}}\frac{\delta\left(\sqrt{-g}\,\mathcal{L}_m\right)}{\delta g^{i j}}\,~~\hspace{-0.0cm}~\text{and}~~~~~ \theta_{i j}=-\frac{2}{\sqrt{-g}}\frac{\delta\left(\sqrt{-g}\,\mathcal{L}_\theta\right)}{\delta g^{i j}},~~~~
\end{eqnarray}
Varying Equation (\ref{1}) with respect to the connection, one obtains
\begin{eqnarray}
\bigtriangledown_i \bigtriangledown_j \left(\sqrt{-g}\,f_Q\,P^k_{\,\,\,\,i j}+H^{k}_{\,\,\,\,i j}\right)=0,~~ \label{10a}
\end{eqnarray}
$H^k_{\,\,\,\,i j}=-\frac{1}{2} \frac{\delta \mathcal{L}_m}{\delta \Gamma^k_{\,\,\,i j}}$, denotes the hyper momentum tensor density. Furthermore, we may deduce the additional restriction over the connection, $\bigtriangledown_i \bigtriangledown_j (H^k_{\,\,\,\,i j})=0$ as follows from Equation (\ref{10a}),
\begin{equation}\label{10}
\bigtriangledown_i \bigtriangledown_j \left(\sqrt{-g}\,f_Q\,P^k_{\,\,\,\,i j}\right)=0.
\end{equation}
{The affine connection has the following form since it is torsionless and curvature less, it can be parametrized more clearly by a collection of functions, such as,
\begin{equation}\label{11}
\Gamma^k_{\,\,\,i j}=\left(\frac{\partial x^k}{\partial\xi^l}\right)\partial_i \partial_j \xi^l.
\end{equation}
An invertible relation is $\xi^k = \xi^k (x^i)$ in this case. As a result, finding a coordinate system that eliminates the $\Gamma^k_{\,\,\,i j}$ connection is always possible i.e. $\Gamma^k_{\,\,\,i j}=0$. The covariant derivative $\nabla_i$ reduces to the partial one $\partial_i$, which is known as the coincident gauge. However, the metric evolution would be modified in any other coordinate system where this affine relationship does not vanish, resulting in an entirely different theory \cite{Dimakis21}.
As a result, we get a gauge coordinate that is coincident, and the nonmetricity Equation (\ref{2}) is reduced to
\begin{equation}\label{12}
Q_{k i j}=\partial_k \,g_{i j},
\end{equation}
As a result, the computation is simplified because the metric is an essential variable. However, in this case, the action no longer remains diffeomorphism invariant, except for standard General Relativity. One can use the covariant formulation of $f(Q)$ gravity to avoid such an issue. Since the affine connection in Equation (\ref{11}) is purely inertial, one could use the covariant formulation by first determining the affine connection in the absence of gravity. We want to discover gravitationally decoupled solutions for $f (Q)$ gravity for compact objects. Thus, we use the standard static spherically symmetric line element of the form, 
\begin{equation}
\label{13}
ds^2=-e^{a(r)}dt^2+e^{b(r)}dr^2+r^2 d\Omega^2,
\end{equation}
where $d\Omega^2=d\theta^2+\sin^2{\theta} \,d\phi^2$. Now we will focus on the anisotropic matter distribution for this study, and the effective energy-momentum tensor $T^{\text{eff}}_{ij}$ may be represented as:
\begin{eqnarray}
\label{14}
&& \hspace{-0.9cm} T^{\text{eff}}_{ij}=\left(\rho^{\text{eff}}+p^{\text{eff}}_t\right)u_{i}\,u_{j}+p^{\text{eff}}_t\,g_{ij}+\left(p^{\text{eff}}_r-p^{\text{eff}}_t\right)v_{i}\,v_{j},
\end{eqnarray}
where $u_{i}$ is the four-velocity vector of the fluid and $\rho^{\text{eff}}$ is the effective density. Apart from $v_{i}$, which is the radial unit space-like vector, $p^{\text{eff}}_r$ is the effective radial pressure in the direction of $u_{i}$, and $p^{\text{eff}}_t$ is the effective tangential pressure orthogonal to $v_{i}$. However, $v_i$ and $u_i$ satisfying the relation $v^iv_i=-1$, $u^iu_i=1$, and $u^iv_i=0$.
Now, the nonmetricity scalar for the metric Equation (\ref{13}) can be calculated as:}
\begin{equation}\label{15}
Q=-\frac{2 e^{-b(r)} \left(1+r a'(r)\right)}{r^2}, 
\end{equation}
where $'$ represents the derivative with respect to the radial
coordinate $r$ only and this expression of $Q$ is based on the zero affine connection. For the anisotropic fluid (\ref{14}), the independent components of the equations of motion (\ref{8}) in $f(Q)$ gravity are given as,
\begin{eqnarray}
&& \hspace{-0.5cm}\rho^{\text{eff}} =\frac{f(Q)}{2}-f_{Q}\Big[Q+\frac{1}{r^2}+\frac{e^{-b}}{r}(a^\prime+b^\prime)\Big],\label{17}\\
&& \hspace{-0.5cm} p^{\text{eff}}_r=-\frac{f(Q)}{2}+f_{Q}\Big[Q+\frac{1}{r^2}\Big],\label{18}\\
&& \hspace{-0.5cm} p^{\text{eff}}_t=-\frac{f(Q)}{2}+f_{Q}\Big[\frac{Q}{2}-e^{-b} \Big\{\frac{a^{\prime \prime}}{2}+\Big(\frac{a^\prime}{4}+\frac{1}{2r}\Big) \nonumber\\&& \hspace{0.5cm} \times(a^\prime-b^\prime)\Big\}\Big],\label{19}\\
&& \hspace{-0.1cm}0=\frac{\text{cot}\theta}{2}\,Q^\prime\,f_{QQ}. \label{20}
\end{eqnarray}
Moving on to the off-diagonal component of Equation (\ref{20}), we can get the answer for the functional form of $f$ as follows: 
\begin{eqnarray}
f_{QQ}=0~\Longrightarrow~f_{Q}=-\alpha_1~\Longrightarrow~f(Q)=-\alpha_1\,Q-\alpha_2 ,~~~~  \label{21}
\end{eqnarray}
where $\alpha_1$ and $\alpha_2$ are constants in this equation. The following exact formulations of equations of motion are obtained by inserting Equations (\ref{15}) and (\ref{21}) into Equations (\ref{17})-(\ref{19}),
\begin{eqnarray}
&& \hspace{-0.4cm}\rho^{\text{eff}} = \frac{1}{2\,r^2} \Big[2\, \alpha_1+2\, e^{-b}\, \alpha_1  \left(r\, b^\prime-1\right)-r^2 \,\alpha_2 \Big],\label{22}\\
&& \hspace{-0.4cm} p^{\text{eff}}_r=\frac{1}{2\,r^2} \Big[-2\, \alpha_1+2\, e^{-b} \,\alpha_1  \left(r\, a^\prime+1\right)+r^2\, \alpha_2\Big],\label{23}\\
&& \hspace{-0.4cm} p^{\text{eff}}_t=\frac{e^{-b}}{4\,r} \Big[2\, e^{b}\, r\, \alpha_2 +\alpha_1\,  \left(2+r a^\prime\,\right) \left(a^\prime-b^\prime\right)\nonumber\\&& \hspace{0.5cm} +2\, r\, \alpha_1 \, a^{\prime \prime}\Big],~~~~\label{24} 
\end{eqnarray}
and the vanishing of covariant derivative of the effective energy-momentum tensor ($T^{\text{eff}}_{ij}$), i.e. $\bigtriangledown^i T^{\text{eff}}_{ij}=0$, provides
\begin{eqnarray}
&&   -\frac{a^\prime}{2}(\rho^{\text{eff}}+p^{\text{eff}}_r)-(p^{\text{eff}}_r)^{\prime}+\frac{2}{r}( p^{\text{eff}}_{t}-p^{\text{eff}}_r)=0.~~~\label{25}
\end{eqnarray}
The Equation (\ref{25}) is called a Tolman-Oppenheimer-Volkoff (TOV) equation in $f(Q)$ gravity with $f(Q)=-\alpha_1 Q-\alpha_2$. Next, we are aiming to apply the gravitational decoupling by means of CGD methodology to solve the system of Equations (\ref{22})-(\ref{24}) for the compact star model. To apply this, we transform the gravitational potentials $a(r)$ and $b(r)$ by introducing two arbitrary deformation functions via decoupling constant $\beta$ as, 
\begin{eqnarray}
 a(r) \longrightarrow \nu(r)+\beta\, \xi(r)~\text{and}~e^{-b(r)} \longrightarrow \mu(r)+\beta\, \psi(r).~~ ~~ \label{eq26}
\end{eqnarray}
where $\xi(r)$ and  $\psi(r)$ are called the geometric deformation functions corresponding to the temporal and radial metric components, respectively. It is usual that when $\beta =0$, the standard $f(Q)$ gravity theory is recovered. Since we are applying here the CGD approach to solve the field equations, then both deformation functions must be non-zero i.e. fix $ \xi(r) \ne 0$ and  $\psi(r)\ne 0$. This indicates that both the radial and temporal components of the metric function are affected. Under the transformations (\ref{eq26}), the decoupled system (\ref{22})-(\ref{24}) gets divide into two subsystems. The first system represents the field equation in purely $f(Q)$ gravity under $T_{i\,j}$ while other system is for the extra source $\theta_{i\,j}$. Furthermore, we suppose the energy-momentum tensor ${T}_{i\,j}$ describes an anisotropic matter distribution given by,
\begin{equation}\label{eq27}
T_{i\,j}=\left(\rho+p_t\right)u_{i}\,u_{j}+p_t\,g_{i\,j}+\left(p_r-p_t\right)v_{i}\,v_{j},
\end{equation}
where $\rho$, $p_{r}$ and $p_{t}$ denote the energy density, radial pressure and tangential pressure, respectively corresponding to seed solution i.e. in $f(Q)$ gravity. Due to this, the effective quantities can be denoted as, 
\begin{eqnarray}
 \rho^{\text{eff}}=\rho+\beta \,\theta^0_0,~~p^{\text{eff}}_r=p_r-\beta\,\theta^1_1,~~p^{\text{eff}}_t=p_t-\beta\,\theta^2_2,~~~~  \label{eq28}
\end{eqnarray}
and respective effective anisotropy,
\begin{eqnarray} 
\Pi^{\text{eff}}=p^{\text{eff}}_t-p^{\text{eff}}_r= \Pi+\Pi_{\theta}, \label{eq29}
\end{eqnarray}
where $~\Pi= p_t-p_r~~\text{and}~~\Pi_\theta= \beta (\theta^1_1-\theta^2_2)\nonumber.$\\ 
It is noted that the effective anisotropy is the sum of two anisotropies corresponding to $T_{i\,j}$ and $\theta_{i\,j}$. The anisotropy $\Pi_{\theta}$ is generated by gravitational decoupling that will change the effective anisotropy but this change will solely depend on the behavior of $\Pi_{\theta}$. Now plugging of Equation (\ref{eq26}) in to the system (\ref{22})-(\ref{24}) which yield the following set of equations of motion depending on the gravitational potentials $\nu$ and $\mu$, (i.e. when $\beta = 0$) as,
\begin{eqnarray}
&&\hspace{-0.8cm}\rho= \frac{\alpha_1 }{r^2}-\frac{\mu \alpha_1 }{r^2}-\frac{\mu^{\prime} \alpha_1 }{r}-\frac{\alpha_2 }{2},\label{eq30}\\
&&\hspace{-0.8cm} p_r=-\frac{\alpha_1 }{r^2}+\frac{\mu \alpha_1 }{r^2}+\frac{\nu^{\prime} \mu \alpha_1 }{r}+\frac{\alpha_2 }{2}, \label{eq31}\\
&&\hspace{-0.8cm} p_t=\frac{\mu^{\prime} \nu^{\prime} \alpha_1 }{4}+\frac{\nu^{\prime \prime} \mu \alpha_1 }{2}+\frac{\nu^{\prime 2} \mu \alpha_1}{4} \nonumber\\&& \hspace{0.5cm} +\frac{\mu^{\prime} \alpha_1 }{2 r}+\frac{\nu^{\prime} \mu \alpha_1 }{2 r}+\frac{\alpha_2 }{2}, \label{eq32}
\end{eqnarray}
and the TOV Equation (\ref{25}) gives, 
\begin{eqnarray}
-\frac{\nu^\prime}{2}(\rho+p_r)-(p_r)^{\prime}+\frac{2}{r}( p_{t}-p_r)=0.~~\label{eq33}
\end{eqnarray}
So, the corresponding solution can be given by the following spacetime,
\begin{equation}
ds^2=-e^{\nu(r)}dt^2+\frac{dr^2}{\mu(r)}+r^2d\theta^2+r^2\text{sin}^2\theta d\phi^2, \label{eq34}
\end{equation}
Moreover, the system of field equations for $\theta$-sector is derived by turning on $\beta$ as,
\begin{eqnarray}
&&\hspace{-0.3cm}\theta^{0}_0=-\alpha_1 \Big(\frac{\psi   }{r^2}+\frac{\psi^\prime }{r}\Big), \label{eq35}\\
&&\hspace{-0.3cm}\theta^1_1=-\alpha_1\Big(\frac{\psi  }{r^2}+\frac{a^{\prime} \psi   }{r}+\frac{\mu\,\xi^{\prime}}{r}\Big), \label{eq36}\\
&&\hspace{-0.3cm}\theta^2_2=-\alpha_1 \Big(\frac{1}{4} \psi^\prime a^{\prime}   +\frac{1}{2} a^{\prime \prime} \psi   +\frac{1}{4} a^{\prime 2} \psi  +\frac{\psi^\prime  }{2 r}+\frac{a^{\prime} \psi  }{2 r}\Big)\nonumber\\&&\hspace{0.5cm}-\alpha_1\Big[\frac{\mu}{4}\,\big(2\,\xi^{\prime\prime}+\alpha_1\,\xi^{{\prime}\,2}+\frac{2\,\xi^{\prime}}{r}+2\,\nu^{\prime}\,\xi^{\prime}\big)+\frac{\mu^{\prime}\,\xi^{\prime}}{4}\Big],~~~ \label{eq37}
\end{eqnarray}
and corresponding conservation is,
\begin{eqnarray}\label{eq38}
-\frac{\nu^{\prime}}{2} (\theta^0_0-\theta^1_1)+ (\theta^1_1)^\prime+\frac{2}{r} (\theta^1_1-\theta^2_2)=\frac{\xi^{\prime}}{2}\,({p_r}+{\rho}).
\end{eqnarray}
On the other hand, the mass function corresponding to both systems is determined by,
\begin{eqnarray}\label{eq39}
 m_{Q}=\frac{1}{2} \int^r_0 \rho(x)\, x^2 dx~~\text{and}~~m_{\phi}= \frac{1}{2}\,\int_0^r \theta^0_0 (x)\, x^2 dx, ~~~
\end{eqnarray}
where $m_{Q}(r)$ and $m_{\phi}(r)$ respectively denotes the mass functions corresponding to the sources $T_{ij}$ and $\theta_{ij}$. Then the interior mass function of minimally deformed space time (\ref{13}) in $f(Q)$ gravity can be given by, 
\begin{eqnarray} \label{eq40}
\hat{m}_Q(r)=m_Q(r)-\frac{\alpha_1\,\beta}{2}\,r\,\psi(r).
\end{eqnarray}

\section{Gravitationally decoupled solution in $f(Q)$ gravity} \label{sec3}
This section contains the gravitationally decoupled solution of the field Equations (\ref{eq30})-(\ref{eq32}) for the strange quark star in $f(Q)$ gravity. To find this, we use the MIT Bag equation of state (EOS) \cite{Chodos:1974} for solving of the seed system (\ref{eq29})-(\ref{eq32}) corresponding to energy-momentum tensor $T_{i\,j}$ that describes the internal structure of the seed model. As it is well known that the MIT Bag model represents degenerated Fermi gas of quarks up ($u$), down ($d$) and strange ($s$) \cite{Chodos:1974,Farhi:1984} that is related with $\rho$ and $p_r$ as,   
\begin{eqnarray} \label{eq41}
{p}_r=\sum_{f} p^{f}-\mathcal{B}_g,~~~~\text{and} ~~\rho=\sum_{f}\rho^{f}+\mathcal{B}_g.
\end{eqnarray}
where $f=u,~d,~s$, while ${p^f}$ and $\rho^f$ denotes the individual pressures and matter densities for each $(u)$, $(d)$ and $(s)$ quark flavor which is neutralized by the total external Bag constant $\mathcal{B}_g$. On the other hand, ${p^f}$ and $\rho^f$ obeys the following relation: $\rho^f=3p^f$. Then the explicit form of the MIT Bag equation of state (EOS) for strange quark stars is, 
\begin{eqnarray} \label{eq42}
p_r=\frac{1}{3}(\rho-4\mathcal{B}_g).
\end{eqnarray} 
Now by plugging of $p_r$ and $\rho$ in above EoS (\ref{eq42}), we find the relation between the metric functions $\mu(r)$ and $\nu(r)$,
\begin{eqnarray} \label{eq43} 
&&\hspace{-1cm} \alpha_1\,(4 \mu+\mu^{\prime} r+3 \nu^{\prime} \mu\, r-4)  +2 r^2 \alpha_2+ 4\,r^2\, \mathcal{B}_g=0, 
\end{eqnarray}
we consider the Tolman IV metric function for $\mu(r)$ to solve the above equation,
\begin{eqnarray}
&& \mu(r)=\frac{1}{1+Lr^2+Nr^4},  \label{eq44}
\end{eqnarray}
where $L$ and $N$ are constants with dimensions $\text{length}^{-2}$ and $\text{length}^{-4}$, respectively. Now integrating of Equations (\ref{eq30}) and (\ref{eq31}) in to Equation (\ref{eq42}) together with above $\mu(r)$, we obtain the solution for $\nu(r)$ as, 
\begin{eqnarray} \label{eq45}
&&\hspace{-0.0cm} \nu(r)=-\frac{1}{18\,\alpha_1}\Big[2 N r^6 (2 \mathcal{B}_g + \alpha_2) + 6 r^2 (2 \mathcal{B}_g - 2 L \alpha_1+ \alpha_2)\nonumber\\&&\hspace{0.5cm}+ 3 r^4 ( L (2 \mathcal{B}_g + \alpha_2)-2 N \alpha_1) -6 \alpha_1 \ln[1 + L r^2 + N r^4]\nonumber\\&&\hspace{0.5cm}+ 18\alpha_1\,C\Big],
\end{eqnarray}
where $C$ is a constant of integration. Using the expression for $\mu(r)$ and $\nu(r)$, we find the expressions for $\rho$, $p_r$, and $p_t$ as,
\begin{eqnarray} \label{eq46}
&&\hspace{-0.5cm}\rho(r)= \frac{1}{2(L r^2+N r^4+1)^2}\Big[L^2(2 \alpha_1 r^2-\alpha_2 r^4)+\alpha_1 L \nonumber\\&& \hspace{0.5cm} \times (4 N r^4+6)-2 \alpha_2 L r^2 (N r^4+1)+N^2 (2 \alpha_1 r^6\nonumber\\&& \hspace{0.5cm} -\alpha_2 r^8)-\alpha_2 -2 N(\alpha_2 r^4-5 \alpha_1 r^2)\Big] ,~~~~~~\\ \label{eq47}
&&\hspace{-0.5cm} p_r(r) =\frac{-1}{6 \left(L r^2+N r^4+1\right)^2}\Big[8 \mathcal{B}_g \left(L r^2+N r^4+1\right)^2+L^2 \nonumber\\&& \hspace{0.5cm} \times ({\alpha_2} r^4-2 {\alpha_1} r^2)+2 L \big(-3 {\alpha_1}+{\alpha_2} N r^6-2 {\alpha_1} N r^4 \nonumber\\&& \hspace{0.5cm} +{\alpha_2} r^2\big)+{\alpha_2}-10 {\alpha_1} N r^2+\{2 {\alpha_2} N r^4+\alpha_2\} N^2 r^8\nonumber\\&& \hspace{0.5cm} -2 {\alpha_1} N^2 r^6\Big],\\ \label{eq48}
&&\hspace{-0.5cm} p_{_t}(r)=\frac{1}{18 {\alpha_1} \left(L r^2+N r^4+1\right)^3}\Big[\{2 {\alpha_2}^2 r^2+8 {\alpha_2}^2 N r^6 \nonumber\\&& \hspace{0.5cm} -37 {\alpha_1}  {\alpha_2}  N r^4+60 {\alpha_1}^2 N r^2-73 {\alpha_1} {\alpha_2} N^2 r^8\}\nonumber\\&& \hspace{0.5cm} -3 {\alpha_1} {\alpha_2}+2 {\alpha_2}^2 N^4 r^{18}+p_{t1}(r)\Big],
\end{eqnarray}
The Equations (\ref{eq46})-(\ref{eq48}) describe the complete spacetime geometry for the seed solution. But still, we need to find the solution of the second system of Equations (\ref{eq35})-(\ref{eq38}) corresponding to the  $\theta$-sector. For this purpose, we propose the well-known approaches, namely: (i) Mimicking of the density constraint i.e. $\rho=\theta^0_0$, and (ii) Mimicking of the radial pressure constraint i.e. $p_r=\theta^1_1$, to solve the second system. The physical motivation of the mimic approaches are described in Ref. \cite{OvallePRD2017}.  

\subsection{\textbf{Mimicking of the density constraint i.e. $\rho=\theta^0_0$}} \label{solA}
To solve the $\theta$-sector, here we mimic the seed energy density $\rho$ to $\theta^0_0$ from Equations (\ref{eq30}) and (\ref{eq35}), we find first order linear differential equation in $\psi(r)$ as, 
\begin{eqnarray}\label{eq49}
\psi^{\prime}+\frac{\psi}{r}+\frac{1}{r}\Big[1-\mu-r\,\mu^{\prime}  -\frac{r^2\,\alpha_2}{2\alpha_1}\Big] =0
\end{eqnarray}
Now we obtain the expression of deformation function $\psi(r)$ after integrating the above equation by using the known potential $\nu(r)$ as,
\begin{eqnarray} \label{eq50}
\psi(r)=\frac{ r^2 \alpha_2\,(1+L r^2+N r^4)-6  \alpha_1 (L r^2+N r^4) }{6 \alpha_1\,(1+L r^2+N r^4)}.
\end{eqnarray}
The arbitrary constant of integration has been taken to be zero to ensure the non-singular nature of $\psi(r)$ at centre. Furthermore, we still need to find the deformation function $\xi(r)$ in order to find the expression for $\theta$-sector. Due to this, we assumed an increasing, positive and non-singular function for $\xi(r)$ as: $\xi(r)=\ln(1+Lr^2+Nr^4$), which yields $a(r)=\nu(r)+\beta\,\xi(r)$ as a increasing function. Hence, the $\theta$-sector components are obtained as, 
\begin{eqnarray}\label{eq51}
&&\hspace{-0.5cm}\theta^0_0(r)= \frac{1}{2(L r^2+N r^4+1)^2}\Big[L^2(2 \alpha_1 r^2-\alpha_2 r^4)+\alpha_1 L\nonumber\\&& \hspace{0.9cm} \times (4 N r^4+6)-2 \alpha_2 L r^2 (N r^4+1)+N^2 (2 \alpha_1 r^6\nonumber\\&& \hspace{0.9cm}-\alpha_2 r^8) -\alpha_2-2 N(\alpha_2 r^4-5 \alpha_1 r^2)\Big] ,~\label{eq52}
\\
&&\hspace{-0.5cm}\theta^1_1(r)=\frac{1}{18 {\alpha_1} \big(L r^2+N r^4+1\big)^2}\Big[4 \mathcal{B}_g \big(L r^3+N r^5+r\big)^2 \nonumber\\&& \hspace{0.9cm} \times \big(\alpha_2-6 \alpha_1 L+\alpha_2 L r^2+\alpha_2 N r^4-6 \alpha_1 N r^2\big) \nonumber\\&& \hspace{0.9cm} -3 \alpha_1 \alpha_2 +2\theta_{11}(r)\Big],
\end{eqnarray}
where $\theta^2_2(r)$ is not written due to a very long-expression. 

\subsection{\textbf{Mimicking of the pressure constraint i.e. $p_r=\theta^1_1$}} \label{solB}
In this mimic constraints approach, we are mimicking the  seed radial pressure $p_r$ with the $\theta^1_1$ from the Equations (\ref{eq31}) and (\ref{eq36}),  we obtain the expression for deformation function $\psi(r)$ as,
\begin{eqnarray} \label{eq53}
\psi(r)=\frac{r^2 \big[\alpha_2+2\theta_{21}(r)+\Psi_2(r)+\alpha_2 N^2 r^8+\Psi_1(r)\big]}{2 \big(L r^2+N r^4+1\big) \big[4 \mathcal{B}_g \big(L r^3+N r^5+r\big)^2+\Psi_1(r)\big]}, ~~
\end{eqnarray}
with 
\begin{eqnarray}
 &&\hspace{-0.3cm}\Psi_1(r)=-\alpha_1\, [4 L^2 r^4+L r^2 \big(6 \beta +8 N r^4+9\big)+4 N^2 r^8\nonumber\\ &&\hspace{0.8cm}+(12 \beta +11) N r^4+3]+2 \alpha_2 \big(L r^3+N r^5+r\big)^2,\nonumber\\  
  &&\hspace{-0.3cm} \Psi_2(r)=L^2 \big(\alpha_2 r^4-2 \alpha_1 r^2\big)+2 L \big(-9 \alpha_1+\alpha_2 N r^6\nonumber\\ &&\hspace{0.8cm}-2 \alpha_1 N r^4+\alpha_2 r^2\big).\nonumber
\end{eqnarray}
Finally using the same deformation function $\xi(r)=\ln(1+Lr^2+Nr^4)$, we find the $\theta$-components for this solution as, 
\begin{widetext}
\begin{eqnarray}\label{eq54}
&&\hspace{-1.5cm} \theta^0_0(r)=\frac{\alpha_1 \big[4 r \big(L r^2+N r^4+1\big) \big(8 \mathcal{B}_g r \big(L+2 N r^2\big) \big(L r^2+N r^4+1\big)+\theta_{30}(r)+\theta_{10}(r)\big)+\theta_{20}(r)\big]}{2 \big(L r^2+N r^4+1\big)^2 \big[r^2\,\theta_{21}(r) -\alpha_1 \big(4 L^2 r^4+\theta_{40}(r)+2 \alpha_2 \big(L r^3+N r^5+r\big)^2\big]^2},~~~~~~~\\\label{eq55}
&&\hspace{-1.5cm} \theta_{1}^{1}(r)=-\frac{\alpha_2+2 \theta_{21}(r)+L^2 \big(\alpha_2 r^4-2 \alpha_1 r^2\big)+\theta_{12}(r)+\alpha_2 N^2 r^8-2 \alpha_1 N^2 r^6+2 \alpha_2 N r^4-10 \alpha_1 N r^2}{6 \big(L r^2+N r^4+1\big)^2}, 
\end{eqnarray}
\end{widetext}
where $\theta_{21}(r)=4 \mathcal{B}_g \big(L r^2+N r^4+1\big)^2$ and the $\theta^2_2(r)$ expression is not written here due to its cumbersome nature. 
\section{Boundary conditions} \label{sec4}
The boundary condition is very important part in the study of the compact star. To study this, we need to find a suitable exterior spacetime or ``vacuum solution"  which should be matched smoothly with interior spacetime at the pressure-free interface i.e. at $r=R$. According to the recent discussion \cite{Wang22}, Schwarzschild Anti-de Sitter spacetime is the most suitable exterior spacetime in $f(Q)$ gravity that can be given as,

\begin{eqnarray}\label{eq56}
ds^2_+ =-\bigg(1-\frac{2{\mathcal{M}}}{r}-{\Lambda \over 3}~r^2\bigg)\,dt^2+\frac{dr^2}{\bigg(1-\frac{2{\mathcal{M}}}{r}-{\Lambda \over 3}~r^2\bigg)} \nonumber\\+r^2 \Big(d\theta^2  +\sin^2\theta\,d\phi^2 \Big),\label{metric1}~~~~~~
\end{eqnarray} 
where $\mathcal{M}$ is the total mass and $\Lambda$ the cosmological constant. Then $\mathcal{M}=\hat{M}_Q/\alpha_1$, and $\Lambda=\alpha_2/2\alpha_1$, where $\hat{m}_Q(R)=\hat{M}_Q$. It is clearly observed that when $\alpha_1=1$ and $\alpha_2=0$, the Schwarzschild Anti-de Sitter spacetime (\ref{eq57}) reduces into Schwarzschild exterior solution. On the other hand, the minimally deformed interior spacetime for the region ($0 \le r \le R$)  is given by,  
\begin{eqnarray}\label{eq57}
 ds^2_{-}= -e^{\nu(r)+\beta\,\xi(r)}\,dt^2+ \big[\mu(r)+\beta\,\psi(r)\big]^{-1} dr^2\nonumber\\+r^2(d\theta^2+\sin^2\theta\,d\phi^2), \label{metric2}~~~~~ 
\end{eqnarray} 
In order to match smoothly the interior and exterior spacetime at the boundary $r=R$, we must use Israel-Darmois matching conditions which state that the solution must satisfy the first and second form at the boundary surface of the star ($r=R$). Mathematically, it can be written as  
\begin{eqnarray}
 \label{eq58}
 e^{a^{-}(r)}|_{r=R}=e^{a^{+}(r)}|_{r=R}~\mbox{and}~e^{b^{-}(r)}|_{r=R}=e^{b^{+}(r)}(r)|_{r=R}, \nonumber \\
\end{eqnarray}
and 
\begin{eqnarray}\label{eq59}
&& \big[G_{i\,\varepsilon}\,r^{\varepsilon}\big]_{\Sigma}\equiv \lim_{r \rightarrow R^{+}} (G_{\epsilon\,\varepsilon})-\lim_{r \rightarrow R^{-}} (G_{\epsilon\,\varepsilon})=0~~\nonumber\\&&\hspace{0.5cm} \Longrightarrow~~~ \big[T^{\text{eff}}_{\epsilon\,\varepsilon}\,r^{\varepsilon}\big]_{\Sigma}=\big[(T_{\epsilon\,\varepsilon}+\beta\,\theta_{\epsilon\,\varepsilon})\,r^{\varepsilon}\big]_{\Sigma}=0,~~~\label{eq63}
\end{eqnarray}
The conditions (\ref{eq58}) and (\ref{eq59}) yields,
\begin{eqnarray}
e^{\nu(R)+\beta \xi(R)} &=& \bigg(1-\frac{2{\mathcal{M}}}{R}-{\Lambda \over 3}~R^2\bigg)~\nonumber\\
\mu(R)+\beta\,\psi(R) &=& \bigg(1-\frac{2{\mathcal{M}}}{R}-{\Lambda \over 3}~R^2\bigg), \label{eq60}\\
p^{\text{eff}}_r(R) &=& p_r(R)+\beta\,\alpha_1\, \,\Big[\psi_{_\Sigma}\Big(\frac{1 }{R^2}+\frac{\nu^{\prime}_{_\Sigma}   }{R}\Big)\nonumber\\&&+\frac{\mu_{_\Sigma}\,\xi^{\prime}_{_\Sigma}}{R} \Big]=0. ~~~~~\label{eq61}
\end{eqnarray}

Using the Equations (\ref{eq60}) and (\ref{eq61}), we have determined the unknown parameters for both solutions such as Bag constant ($\mathcal{B}_g$), mass ($\mathcal{M}$) and arbitrary constant ($C$) to calculate the numerical value. 
\section{Physical Analysis and Astrophysical implications } \label{sec5}

\begin{figure*}[!htb]
    \centering
    \includegraphics[width=4.5cm,height=4cm]{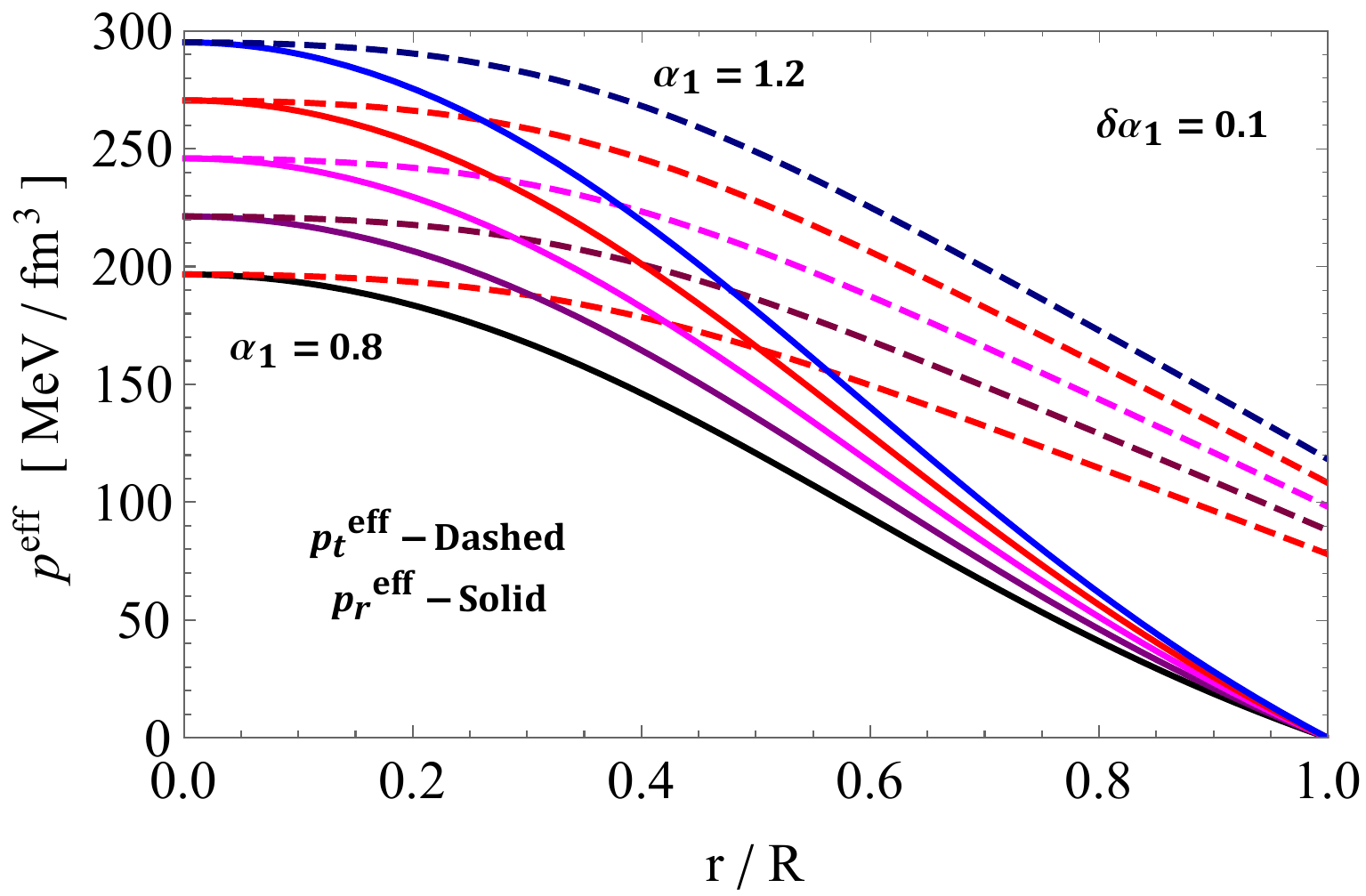}~ \includegraphics[width=4.5cm,height=4cm]{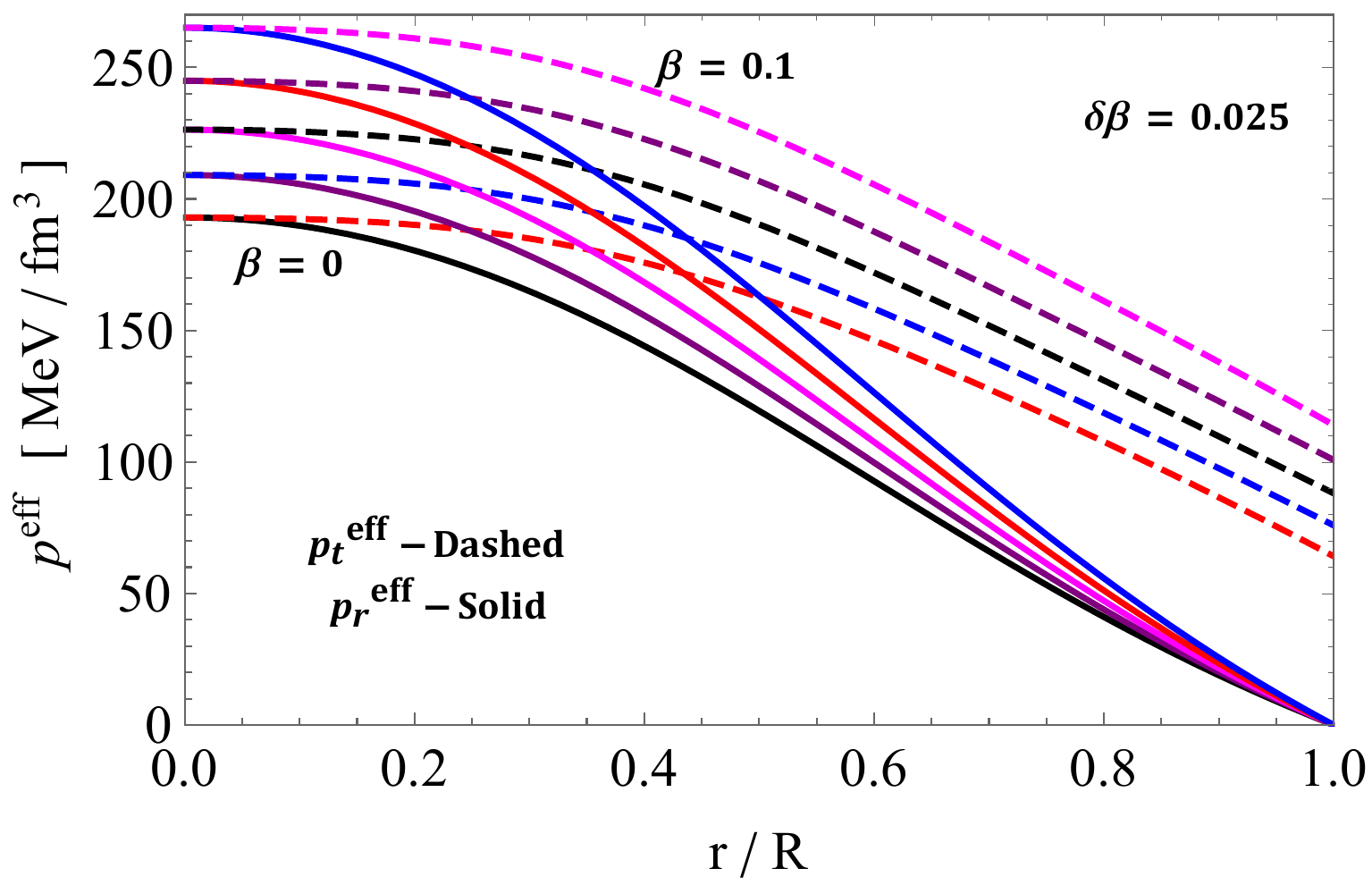}~
     \includegraphics[width=4.5cm,height=4cm]{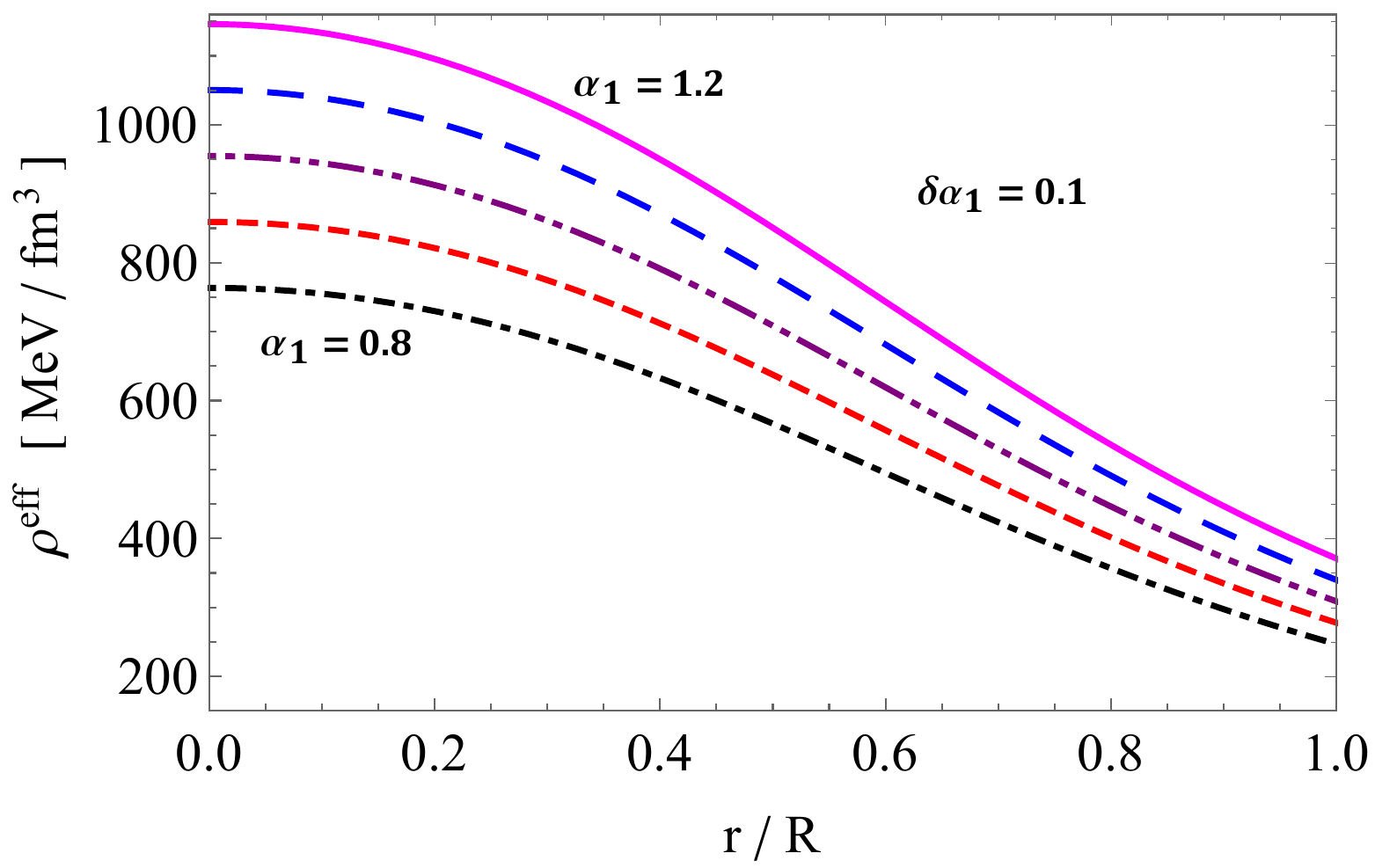}~  \includegraphics[width=4.5cm,height=4cm]{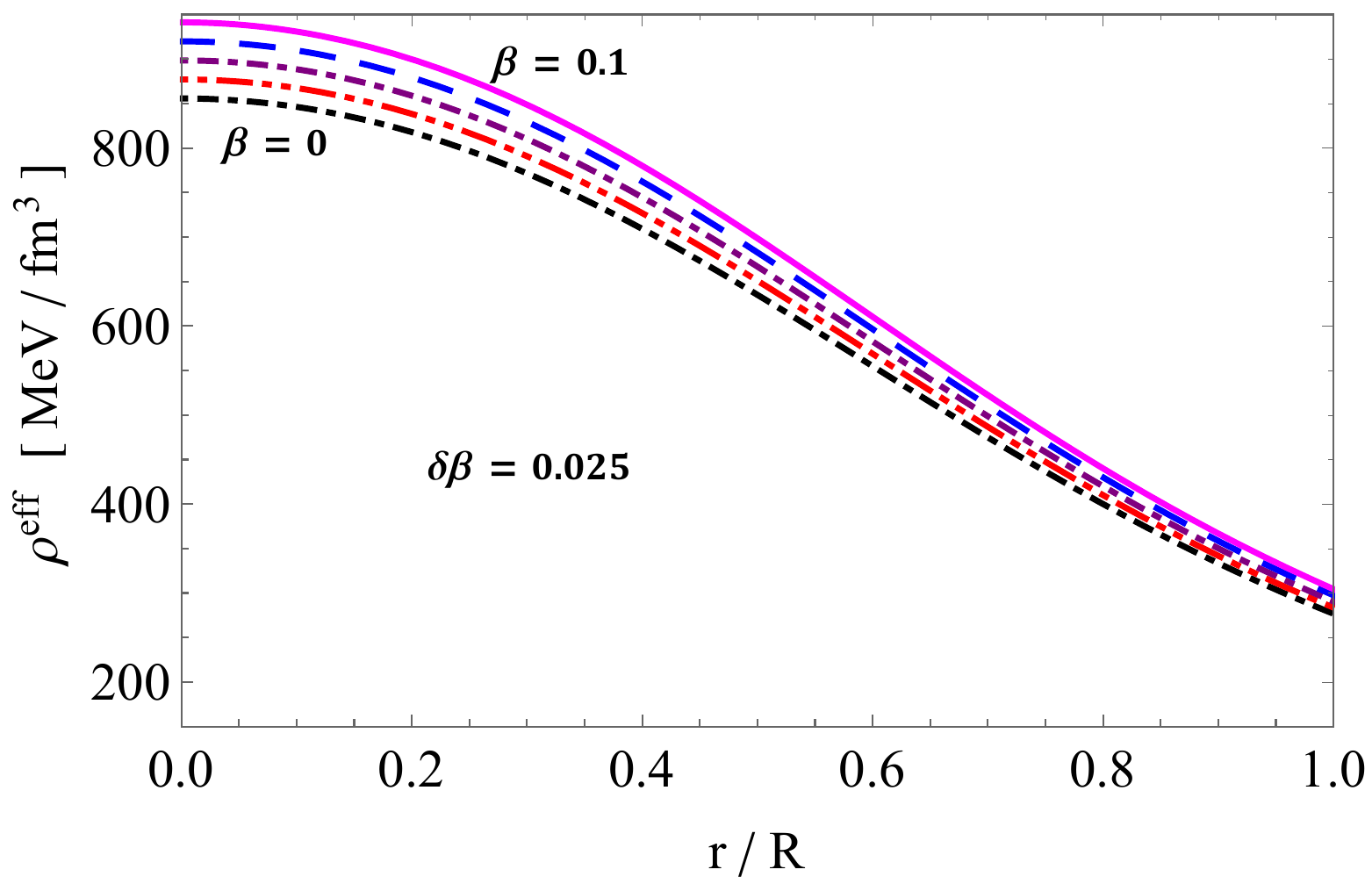}
    \caption{The behavior of radial pressure ($p^{\text{eff}}_r$), tangential pressure ($p^{\text{eff}}_{t}$) and energy density ($\rho^{\text{eff}}$) with respect to $r/R$ for different $\alpha_1$ with fixed $\beta=0.06$ and different $\beta$ with fixed $\alpha_1=0.95$ for the $\rho=\theta^0_0$ solution. 
 We set the numerical values $~L = 0.01/km^2,~N =5\times 10^{-5} /km^4,~R =11.5 \,km$ with $\alpha_2=0.0001$ for plotting of these graphs.}
    \label{fig1}
\end{figure*}

\begin{figure*}[!htb]
    \centering
    \includegraphics[width=4.5cm,height=4cm]{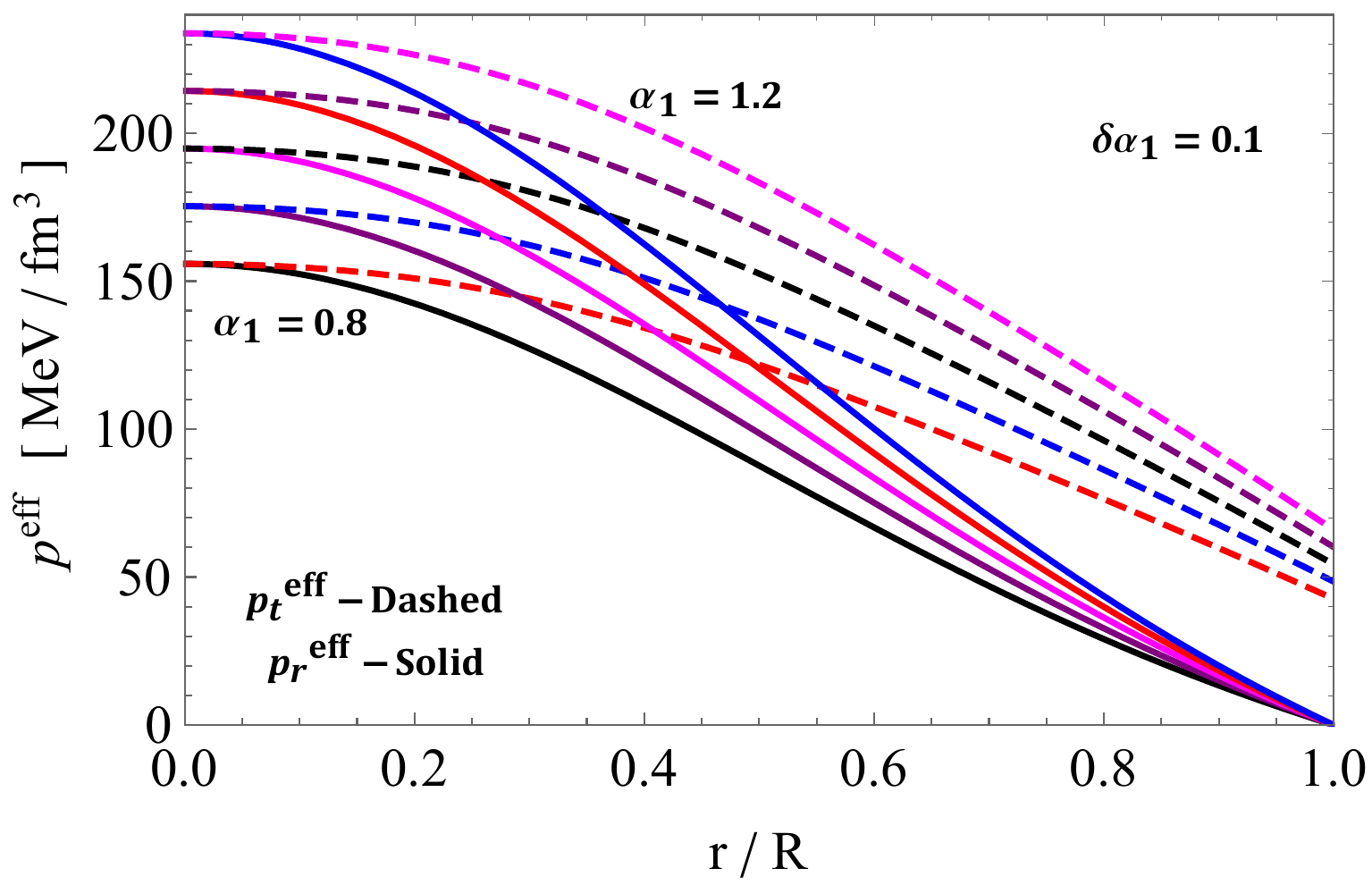}~ 
    \includegraphics[width=4.5cm,height=4cm]{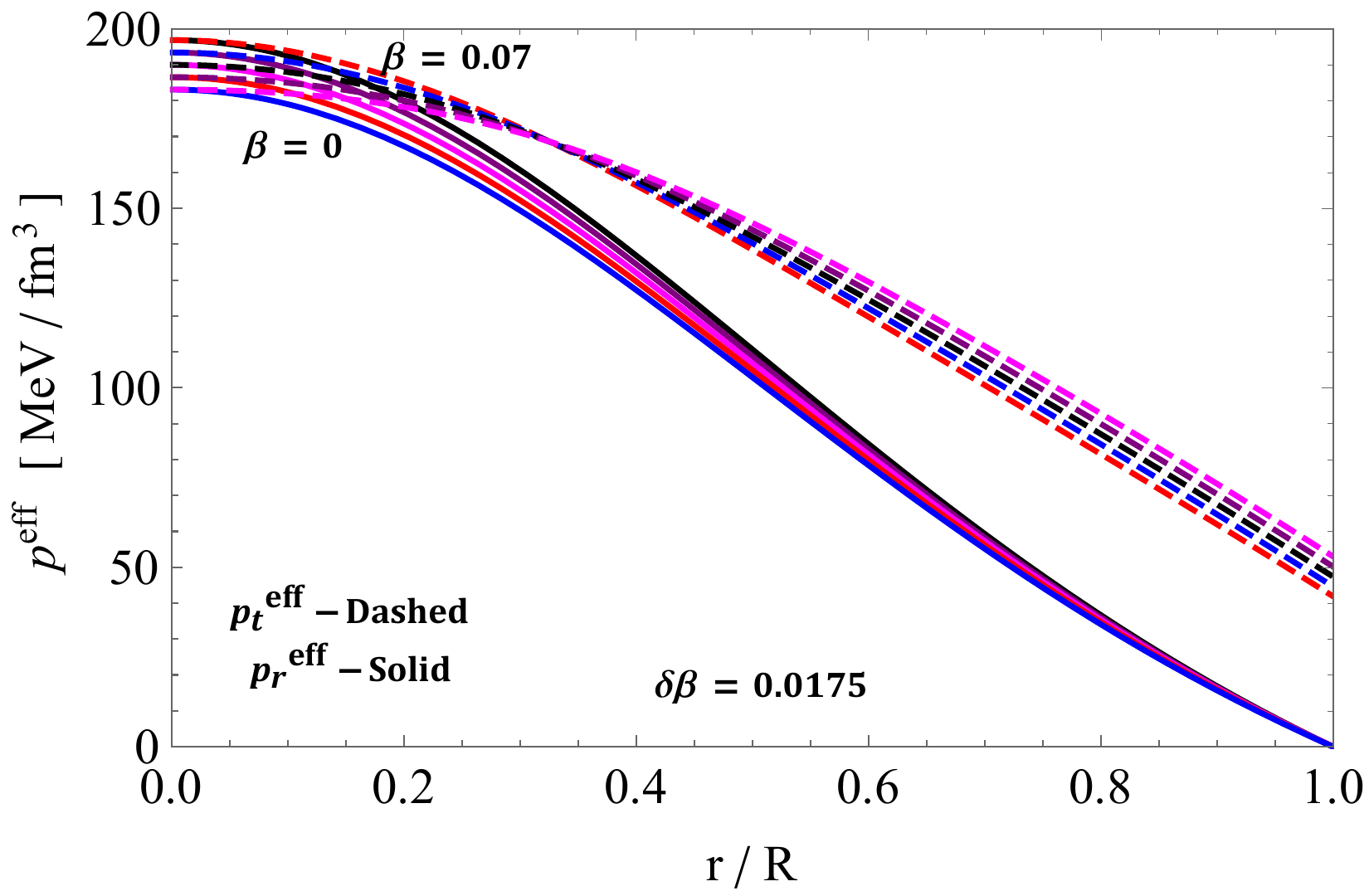}~ \includegraphics[width=4.5cm,height=4cm]{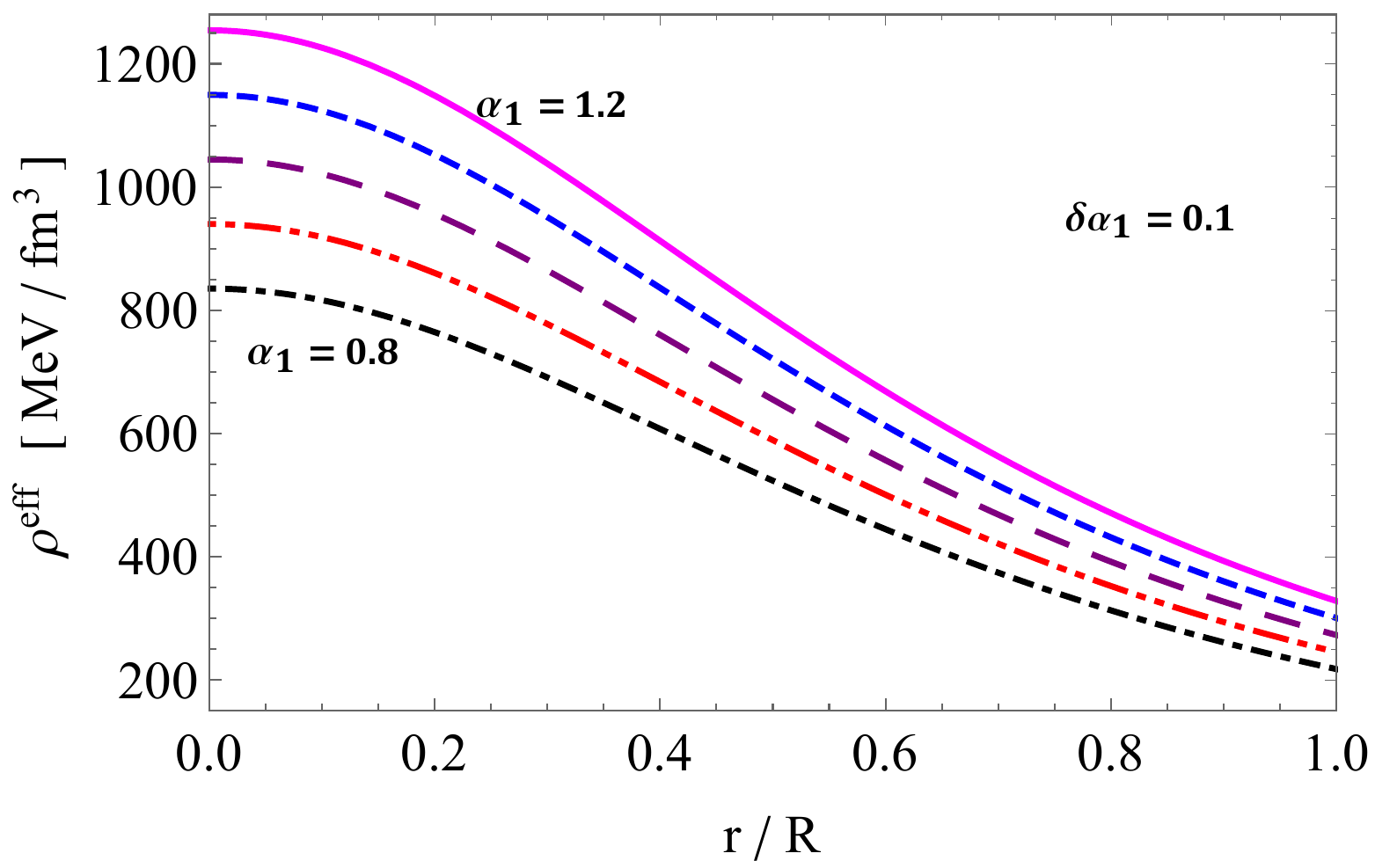}~ 
    \includegraphics[width=4.5cm,height=4cm]{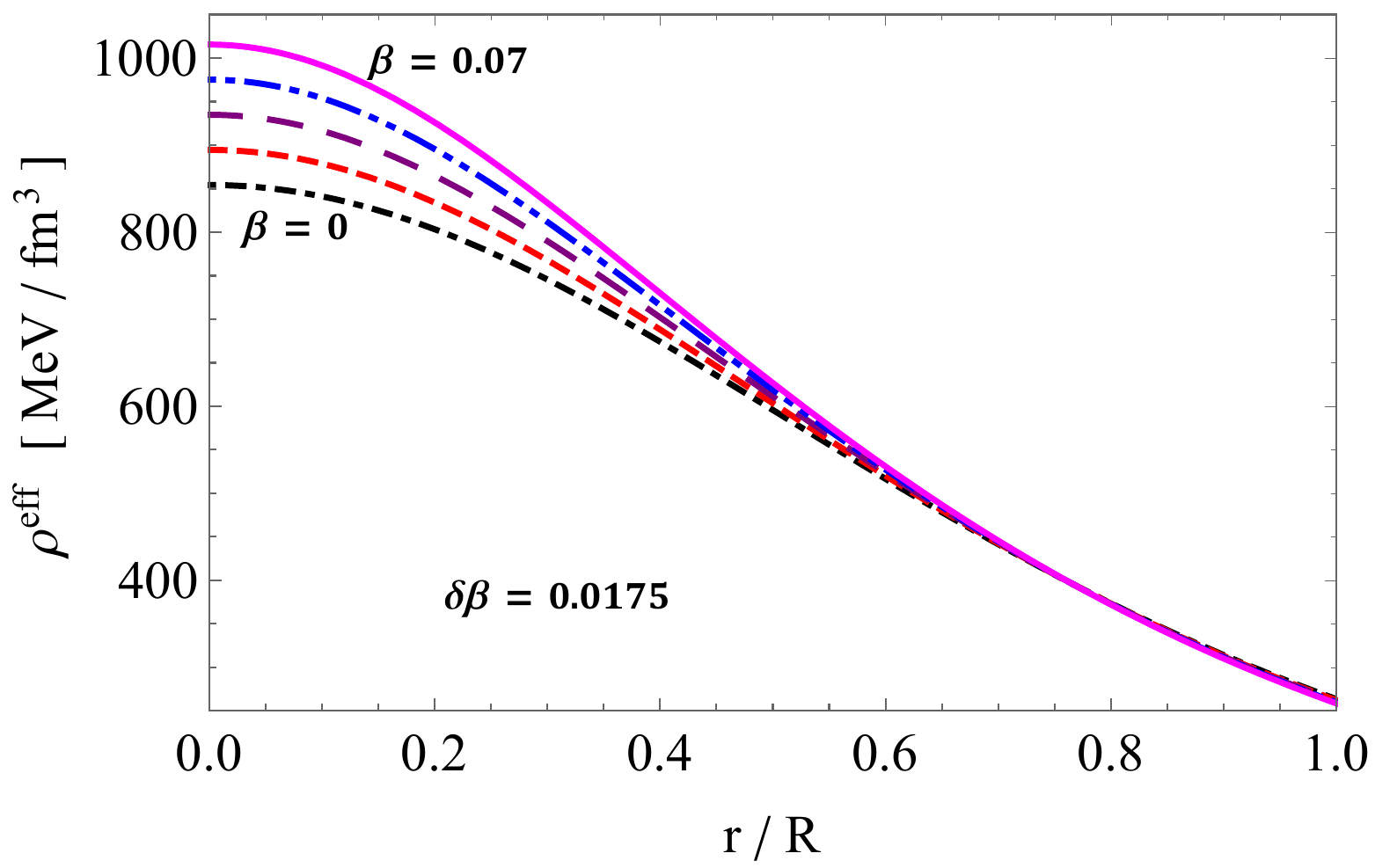}~ 
    \caption{The behavior of radial pressure ($p^{\text{eff}}_r$), tangential pressure ($p^{\text{eff}}_{t}$) and energy density ($\rho^{\text{eff}}$) with respect to $r/R$ for different $\alpha_1$ with fixed $\beta=0.06$ and different $\beta$ with fixed $\alpha_1=0.95$ for the $p_r=\theta^1_1$ solution. 
 We set the numerical values $~L =0.01/km^2,~N = 3\times 10^{-5}/km^4,~R =10.5 \,km$ with $\alpha_2=0.0002$ for plotting of these graphs.}
    \label{fig2}
\end{figure*}

\begin{figure*}[!htb]
    \centering
    \includegraphics[width=4.5cm,height=4cm]{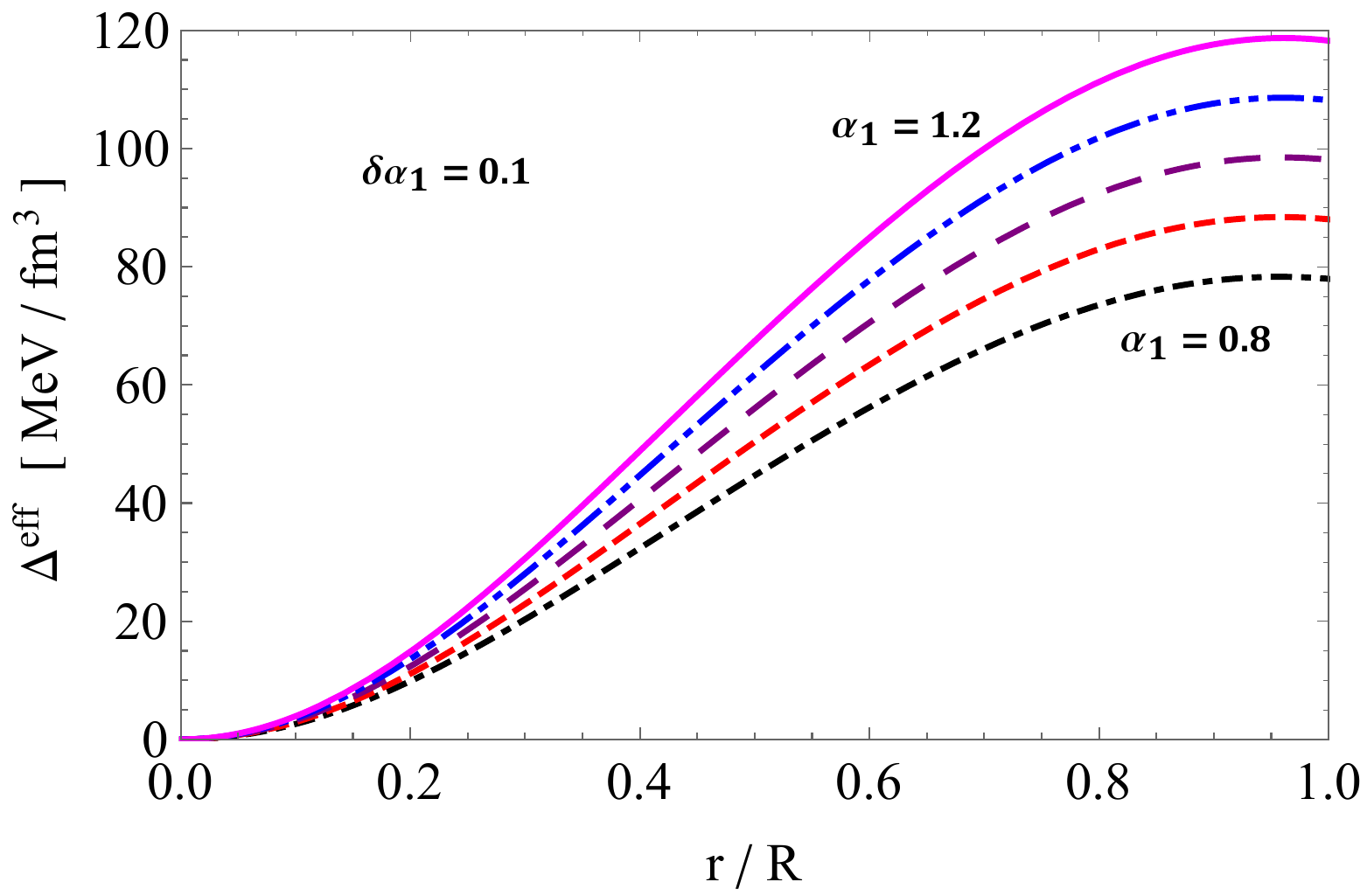}~ \includegraphics[width=4.5cm,height=4cm]{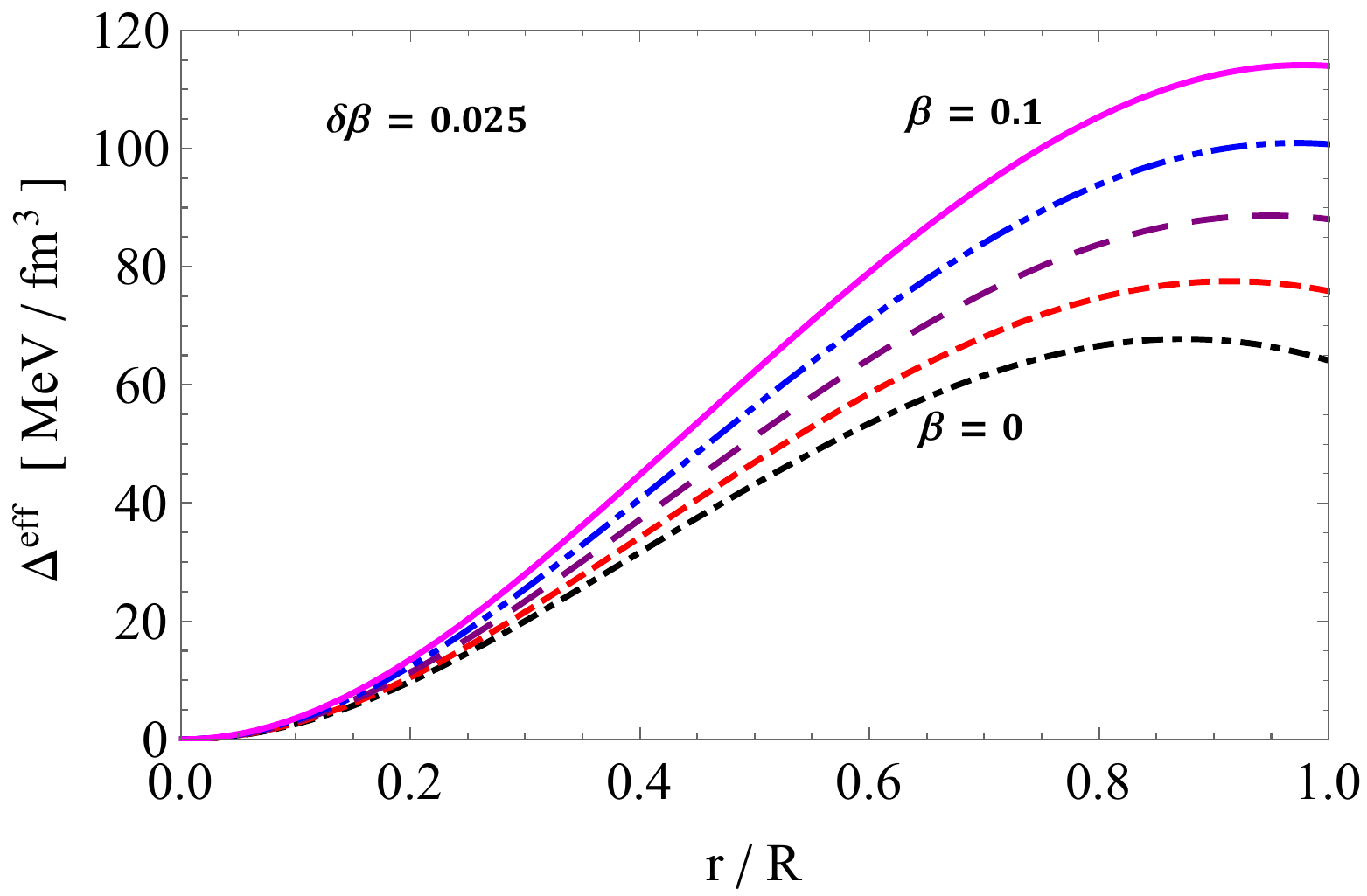}~
     \includegraphics[width=4.5cm,height=4cm]{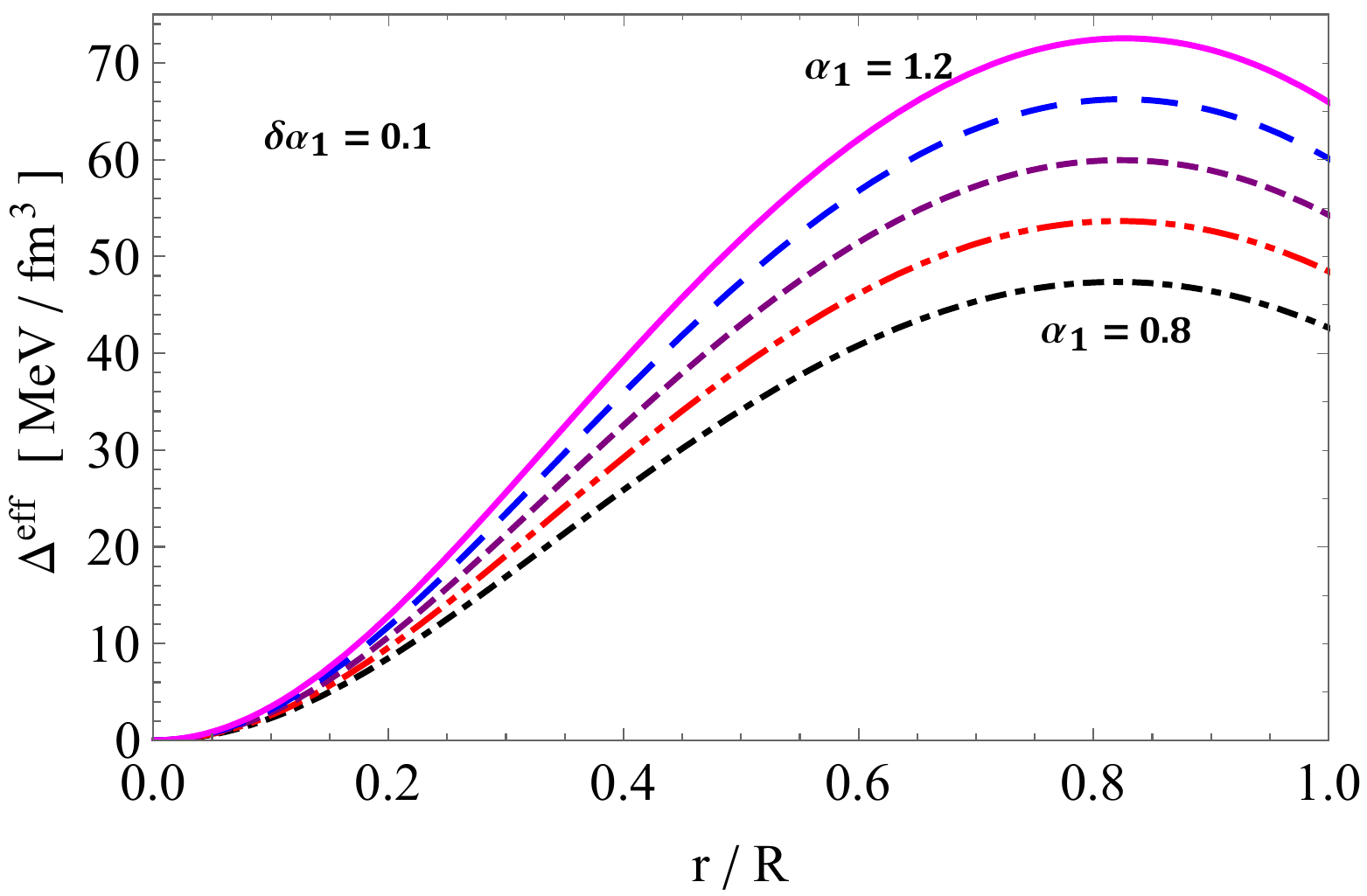}~  \includegraphics[width=4.5cm,height=4cm]{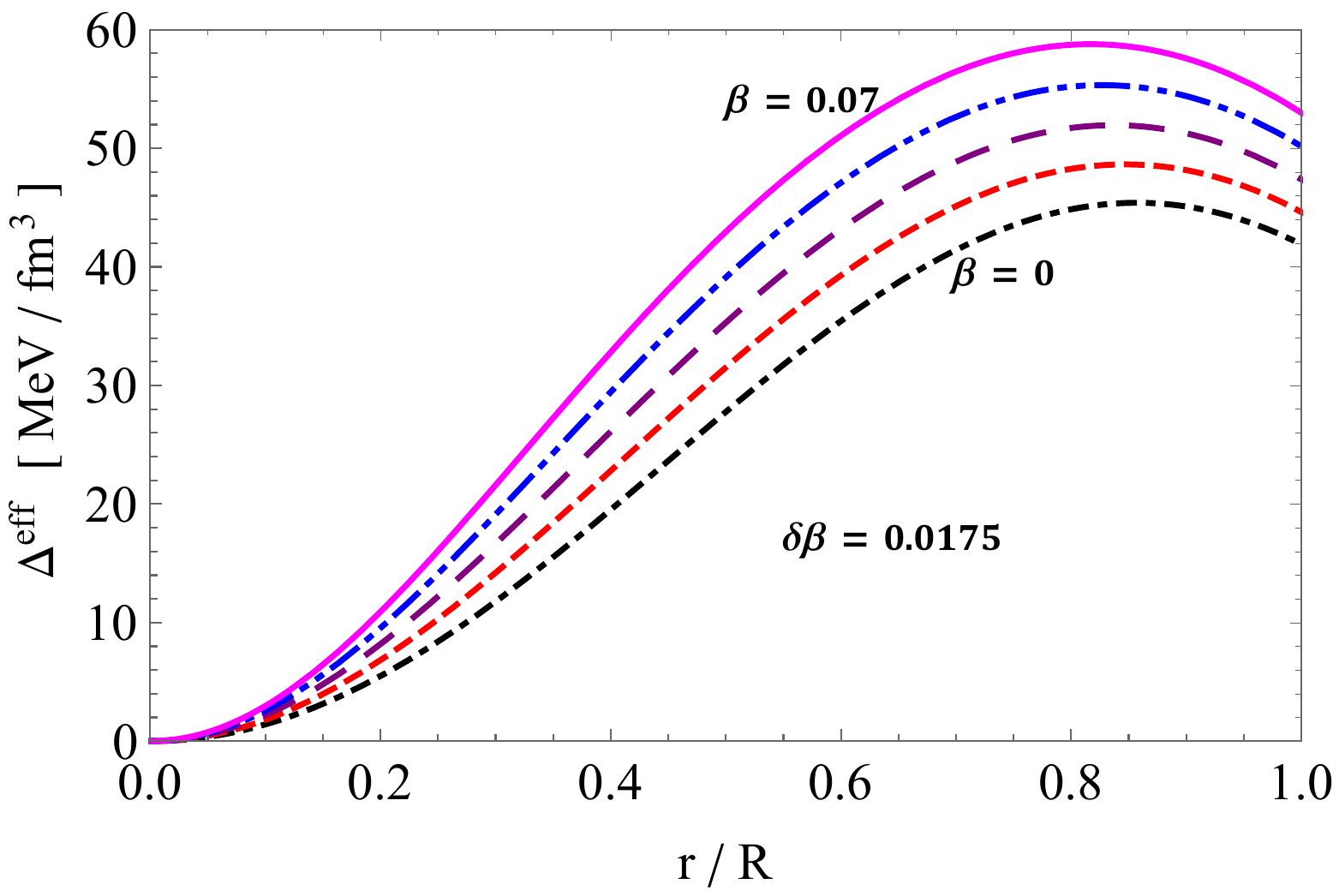}
    \caption{The behavior of anisotropy ($\Delta^{\text{eff}}$) with respect to $r/R$ for different $\alpha_1$ with fixed $\beta=0.06$ and different $\beta$ with fixed $\alpha_1=0.95$ for the $\rho=\theta^0_0$ solution (first two panels) $p_r=\theta^1_1$ solution (last two panel). We set same numerical values as used in Figure \ref{fig2}}
    \label{fig7}
\end{figure*}

\begin{figure*}[!htb]
    \centering
    \includegraphics[width=4.5cm,height=4cm]{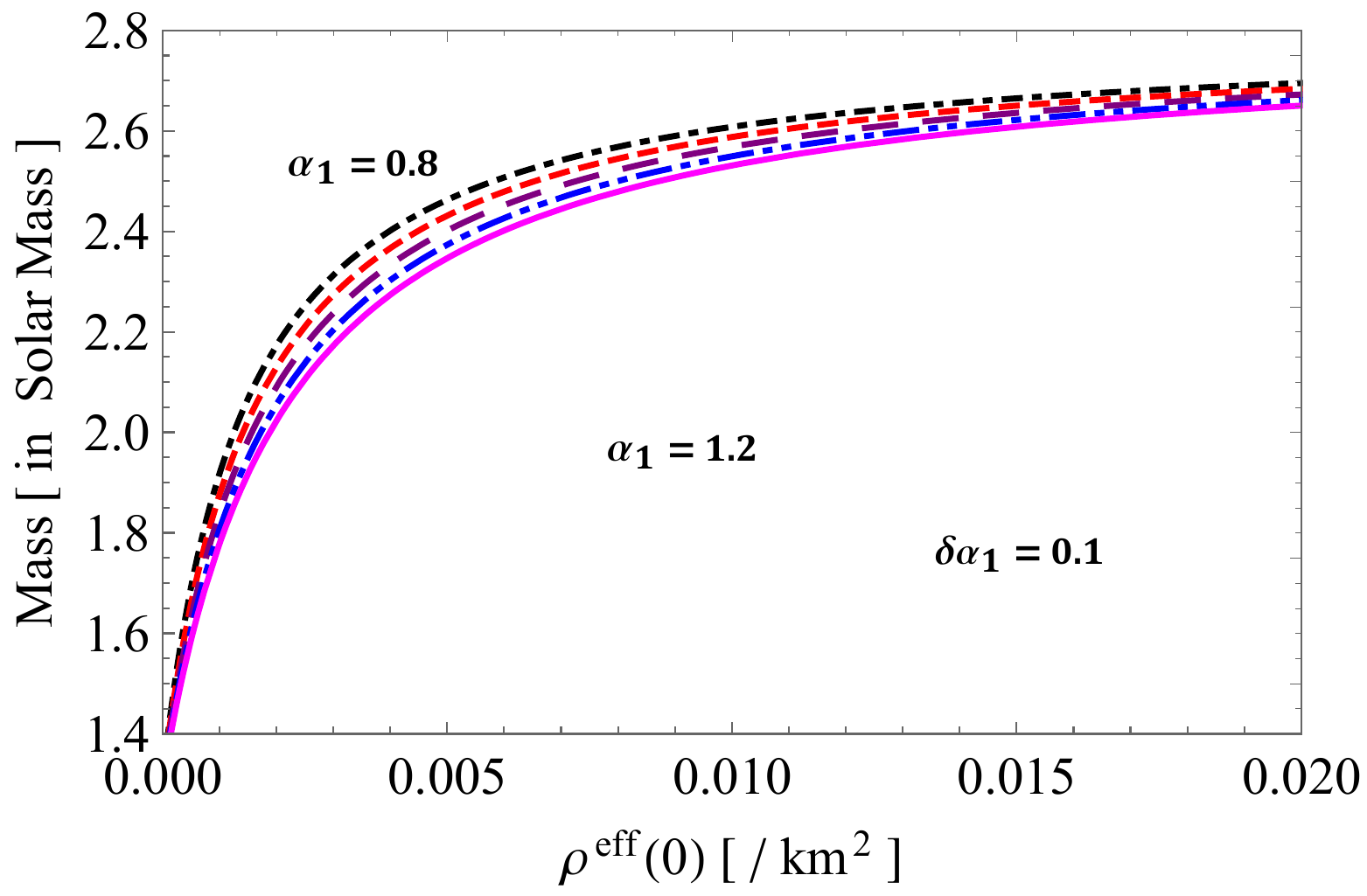}~ \includegraphics[width=4.5cm,height=4cm]{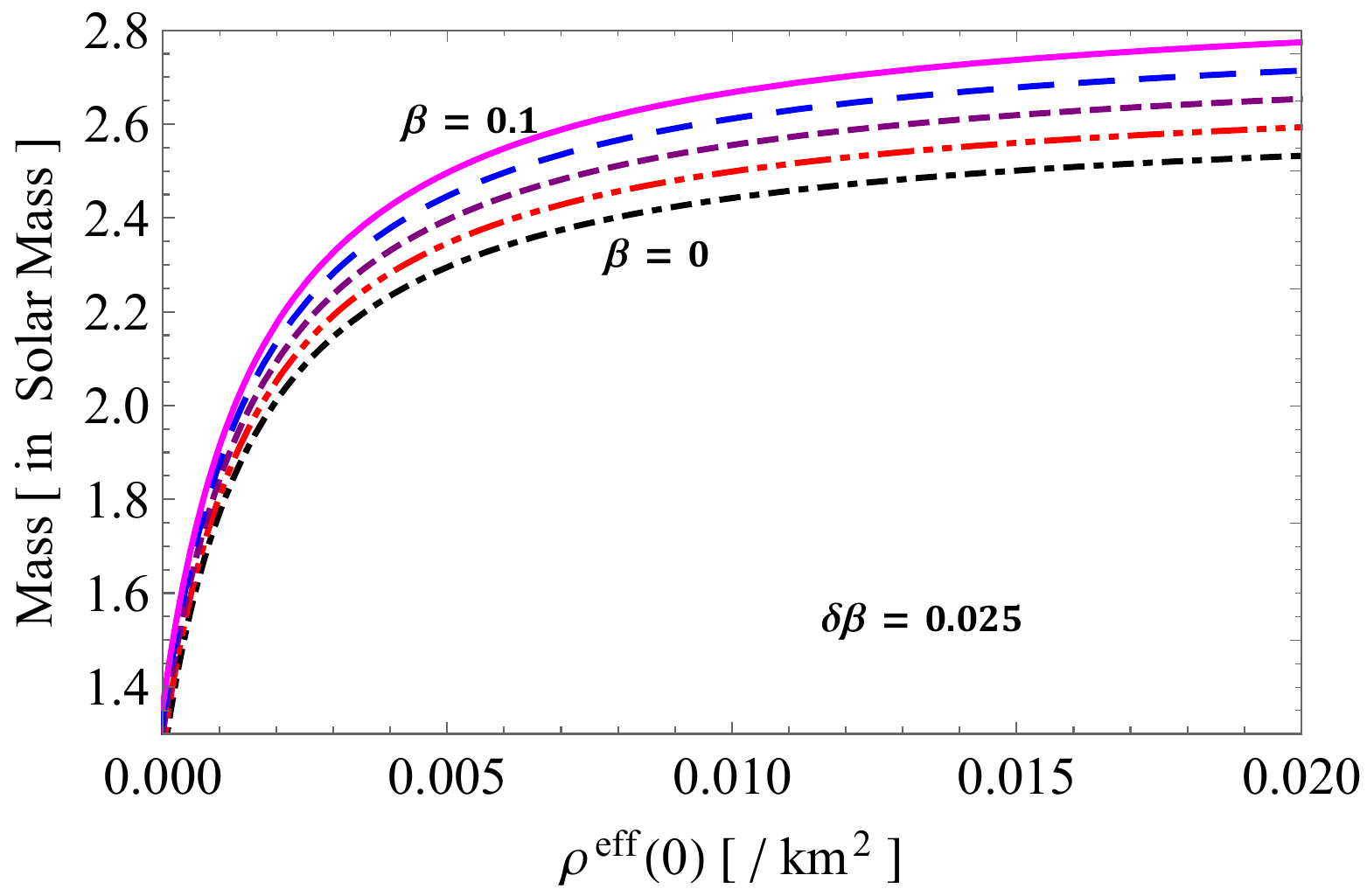}~
     \includegraphics[width=4.5cm,height=4cm]{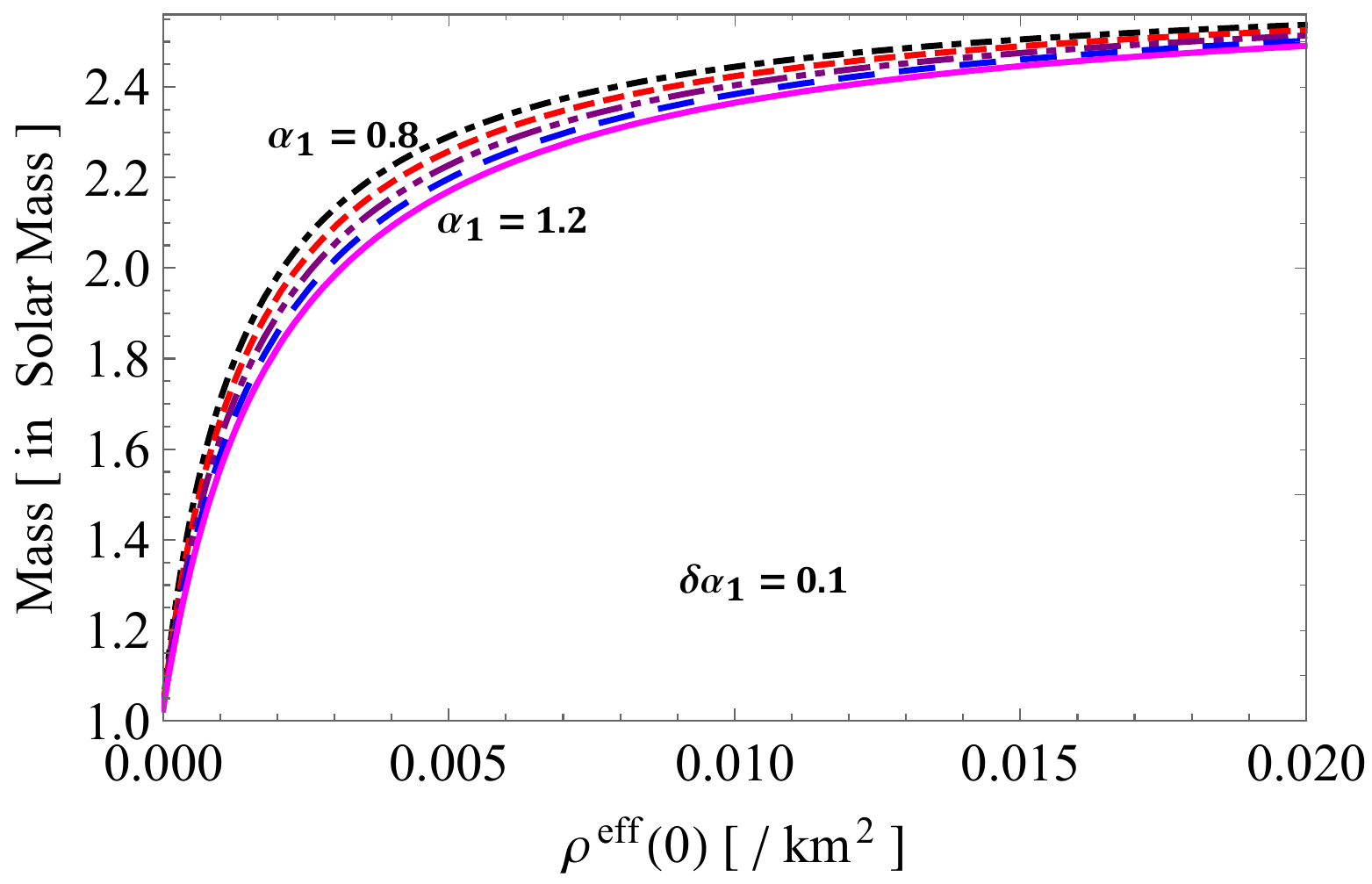}~  \includegraphics[width=4.5cm,height=4cm]{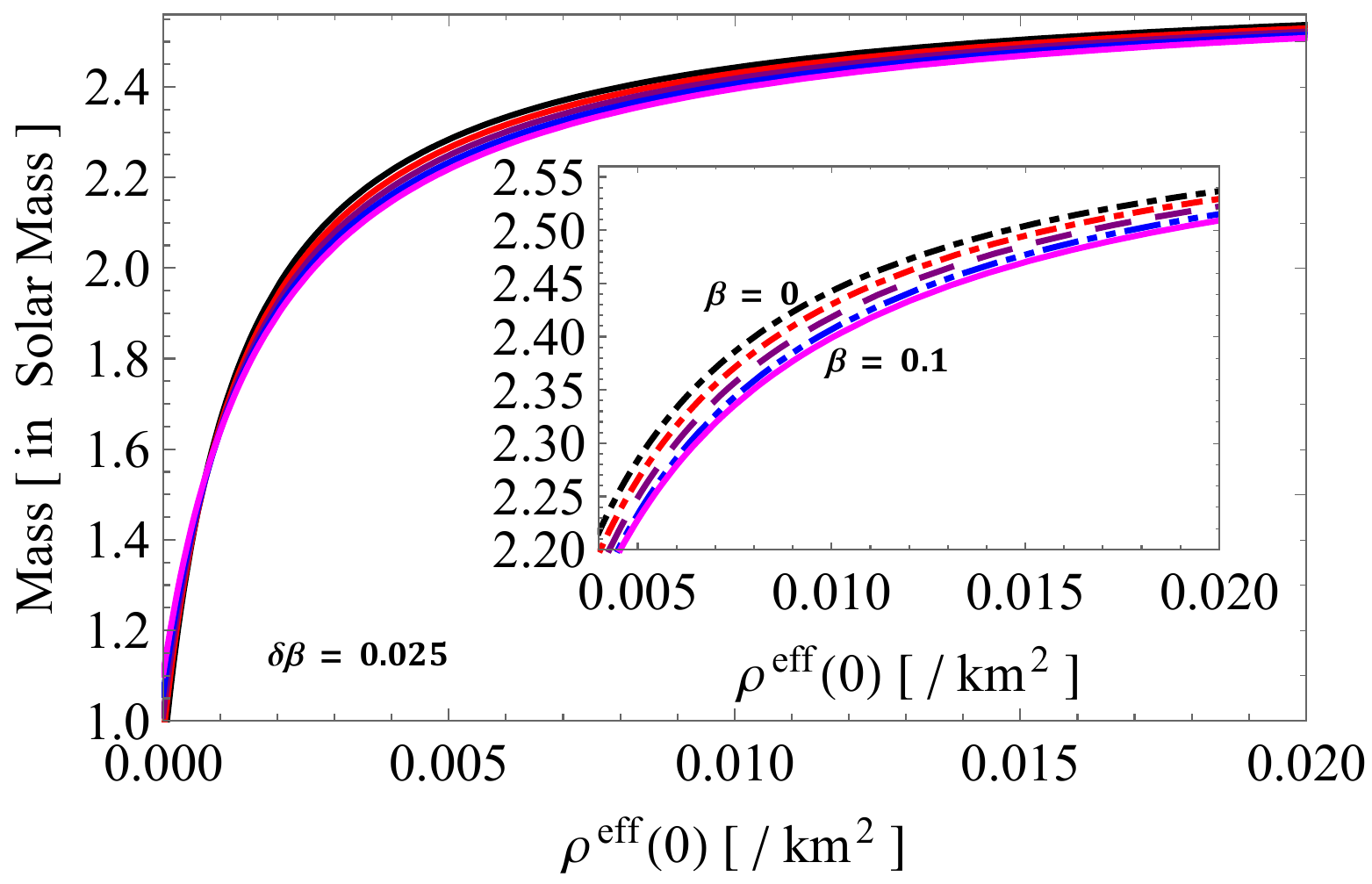}
    \caption{Mass versus central density for different $\alpha_1$ and  $\beta$ for the solution $\theta_0^0=\rho$-(left two panels) and $\theta_1^1=p_r$-(right two panels. We set same numerical values as used in Figure \ref{fig1} and  Figure \ref{fig2}.}
    \label{fig3}
\end{figure*}

In this section, we investigate the behavior of the physical parameters of anisotropic deformed strange star model structures and the model parameters as a function of physical conditions to ensure that the model can accurately explain the stellar structure of strange anisotropic stars. We also chose the physical properties of stars, such as energy density radial and tangential pressures sound speeds at the massive star. In Figure \ref{fig1}-\ref{fig6}, we have also drawn the graphs of the various physical quantities and the physical condition.

Figure \ref{fig1} shows the graphical depiction of energy density for four compact star models under $f(Q)$ gravity. It is positive throughout the set-up of all parameters for the same configuration (see Figure \ref{fig2} ). The behavior of the pressure and density profiles for the mimicking of the component to seed radial pressure is shown in Figure \ref{fig3}. The behavior of the radial and tangential pressures for variable and fixed pressures is shown in the top left panel. The radial and tangential pressures both decrease as they approach the boundary for the $\rho=\theta_{0}^{0}$ and $p_r=\theta_{1}^{1}$ solutions, as seen in Figure \ref{fig1} and Figure \ref{fig2}, respectively.

The behavior of the radial pressure as a function of the scaled radial coordinate $r/R$, is shown in Figure \ref{fig1} and Figure \ref{fig2}. The radial pressure is a monotonically declining function that eliminates over a finite radius, $r = R$, which defines the boundary of the compact object. The effective pressure is decreased much more when the CGD coupling constant is raised. Figure \ref{fig1}. depicts the behavior of the tangential pressure. When the non metricity parameter $(\alpha_1)$ and the decoupling constant ($\beta$) are increased, the behavior is virtually identical to that seen for $p^{\text{eff}}_r$. The overall behavior of energy density as a function of the scaling radial coordinate is studied. Higher core densities result has been observed from an increase in the magnitude of the nonmetricity parameter without CGD effects. The density profile in the bottom left panel is a decreasing function of the scaled radial coordinate-higher energy densities within the core resulting from an increase in the size of the nonmetricity parameter. 

The variability of the anisotropy parameter for the two sectors has been shown in Figure \ref{fig7}. The anisotropy ($\Delta^{\text{eff}}$), is positive at each interior point of the stellar configuration. Due to anisotropy, the tangential pressure dominates the radial pressure i.e. $p_t > p_r$, resulting in a repulsive force. The object is stabilized by this repulsive force against inward gravitational forces. Furthermore, we observe that the highest amount of anisotropy created in the $\theta^0_0 =\rho $ models is 70\% larger than its equivalents in the $\theta^1_1 = p_r$ models. The star is stabilized by this repulsive force, which helps it resist the internally directed gravitational effect. The nonmetricity parameter and the decoupling constant can also be used to adjust the degree of anisotropy.  An increase in anisotropy is observed when $\alpha_1$ is fixed and $\beta$ increased. Overall, when one gets closer to the surface layers of the compact object, the anisotropy increases the most. In addition to the explanation above, the Harrison-Zeldovich-Novikov (HZN) stability criteria is used to investigate the stability of the strange star model. The solutions of the model corresponding to the $\theta_0^0=\rho$ and $\theta_1^1=p_r$ have been discussed. For central densities smaller than $0.0025 km^{-3}$, Figure \ref{fig3} exhibits a similar tendency. We adjusted the nonmetricity parameter while keeping it constant. We should also remark that $dM/d\rho^{\text{eff}}$ is smaller than its equivalent in Figure \ref{fig3}. As a result, we may deduce that the nonmetricity parameter has a more significant impact on fluid configuration stabilization than the decoupling constant.


\subsection{Measurements of the mass-radius relation of observed compact objects via $M-R$ curves} 

\begin{figure*}[!htb]
    \centering
    \includegraphics[width=4.5cm,height=4.1cm]{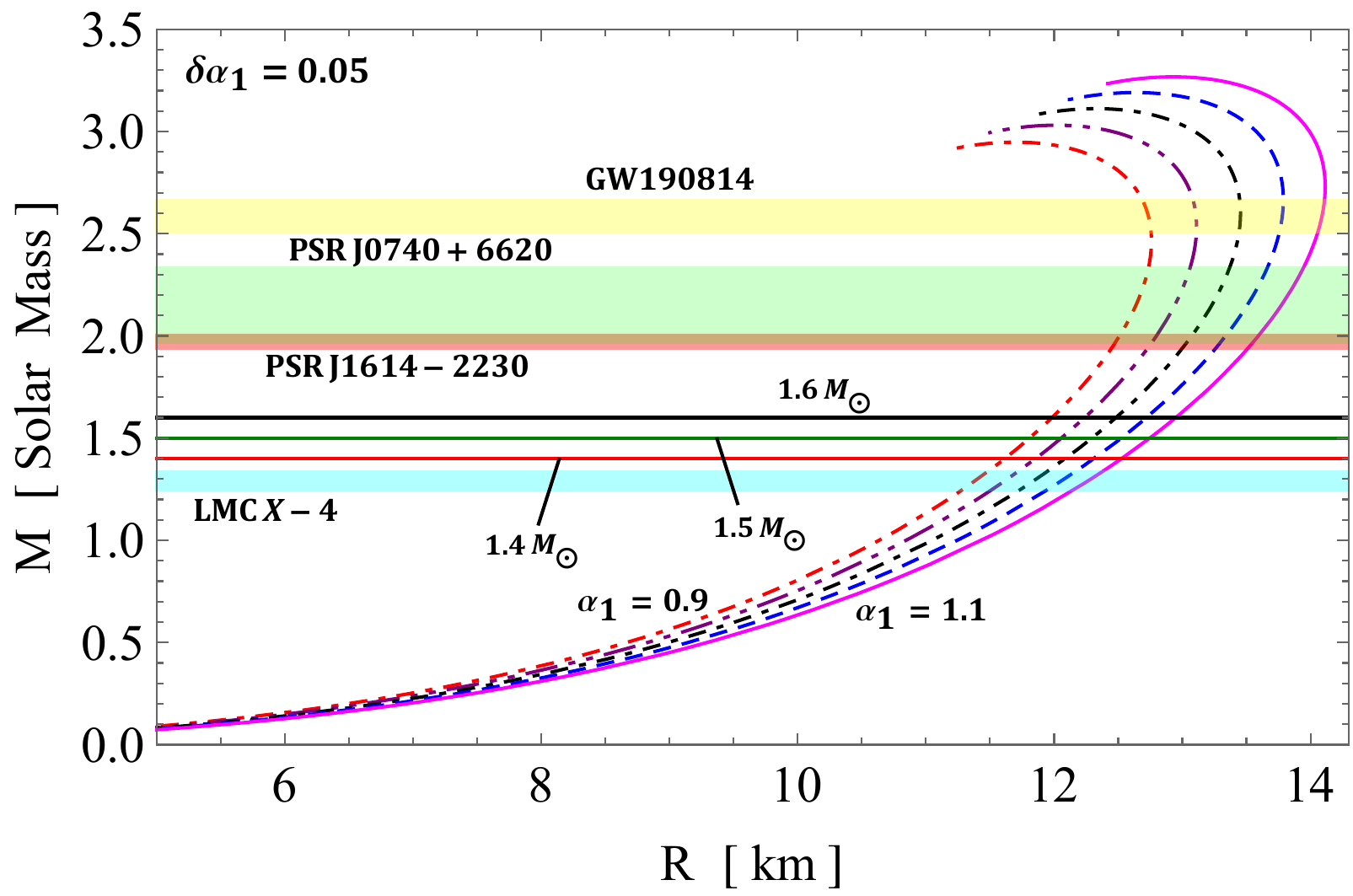}~\includegraphics[width=4.5cm,height=4cm]{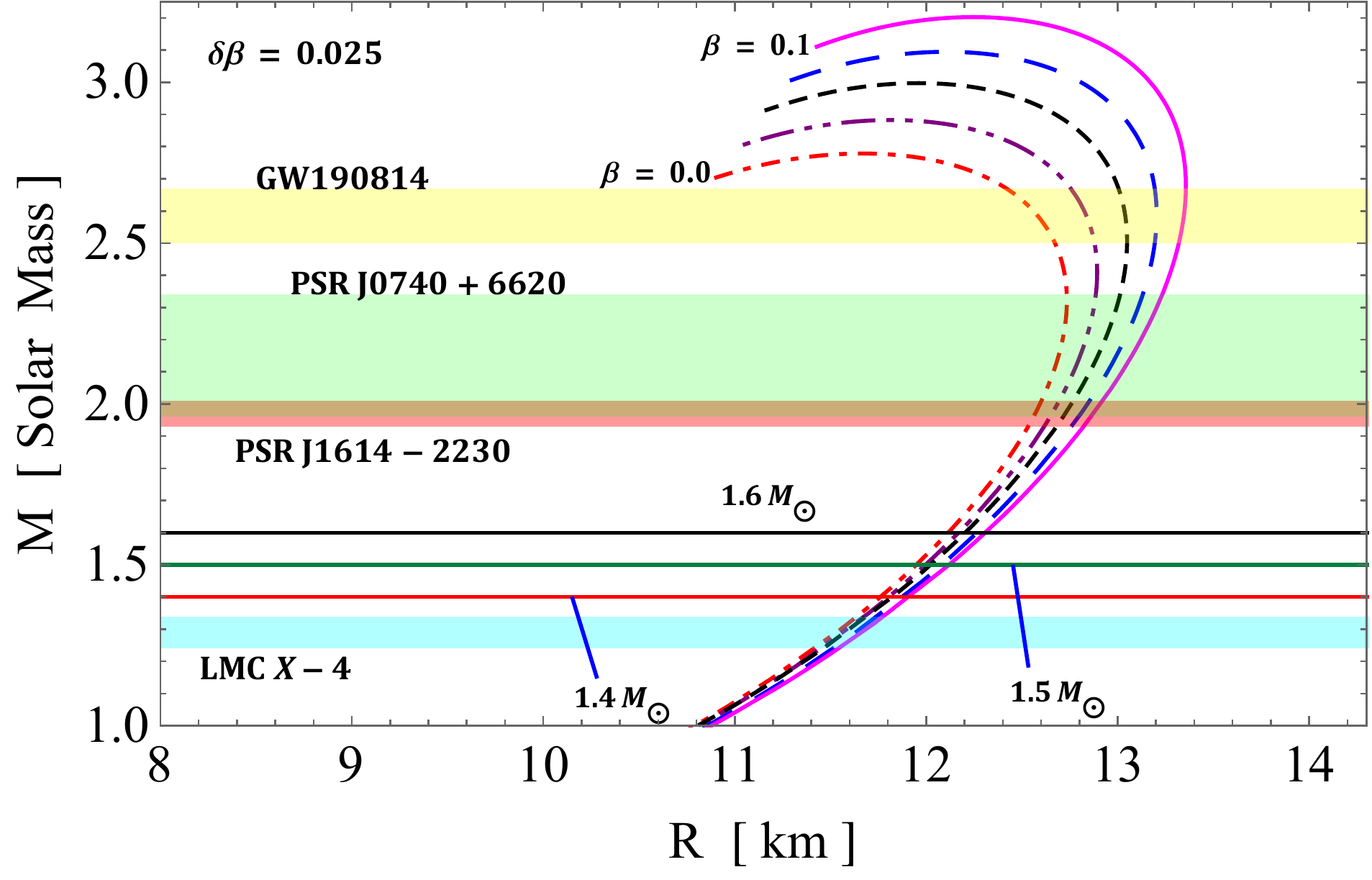}~\includegraphics[width=4.5cm,height=4cm]{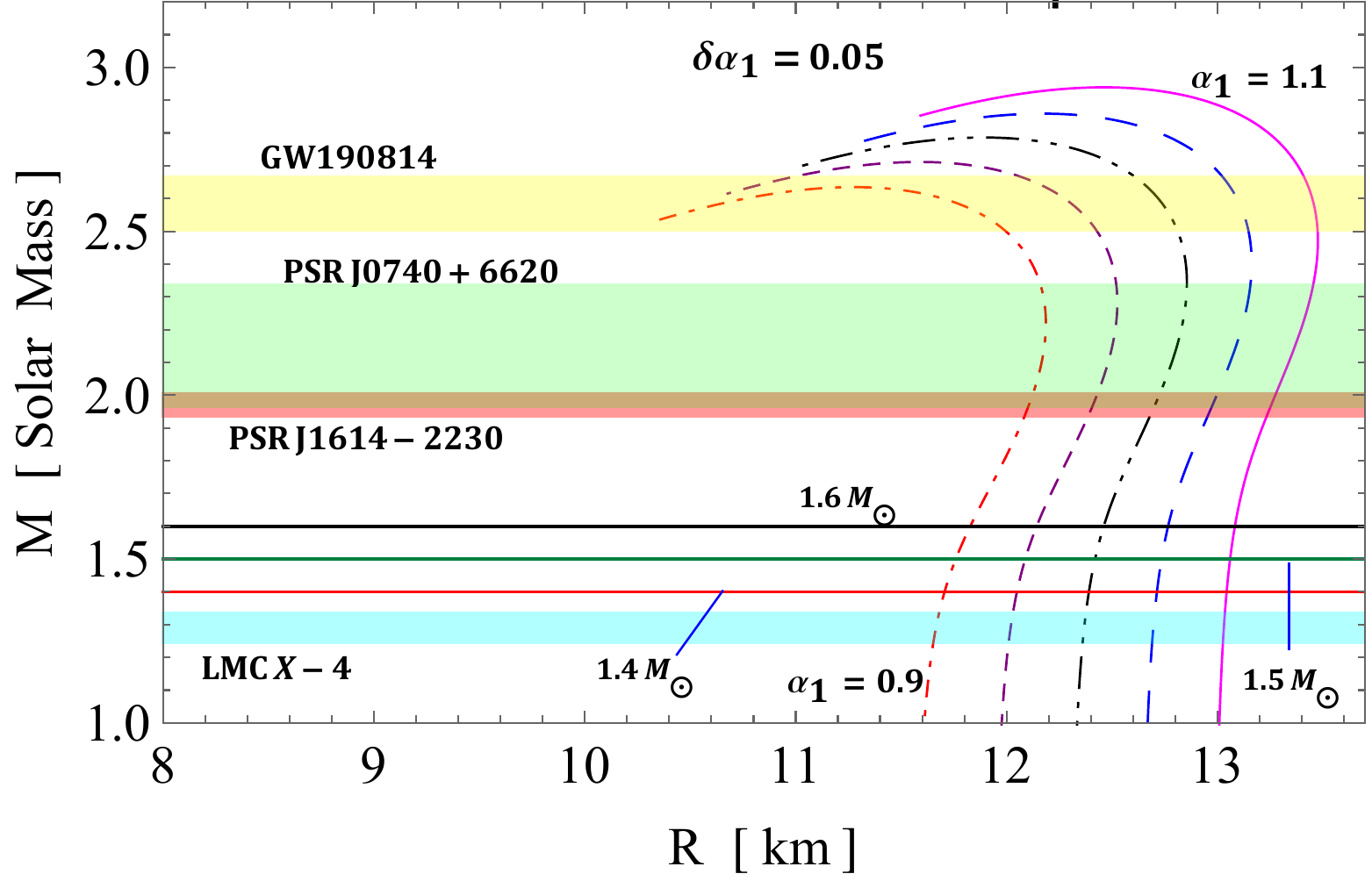}~\includegraphics[width=4.5cm,height=4cm]{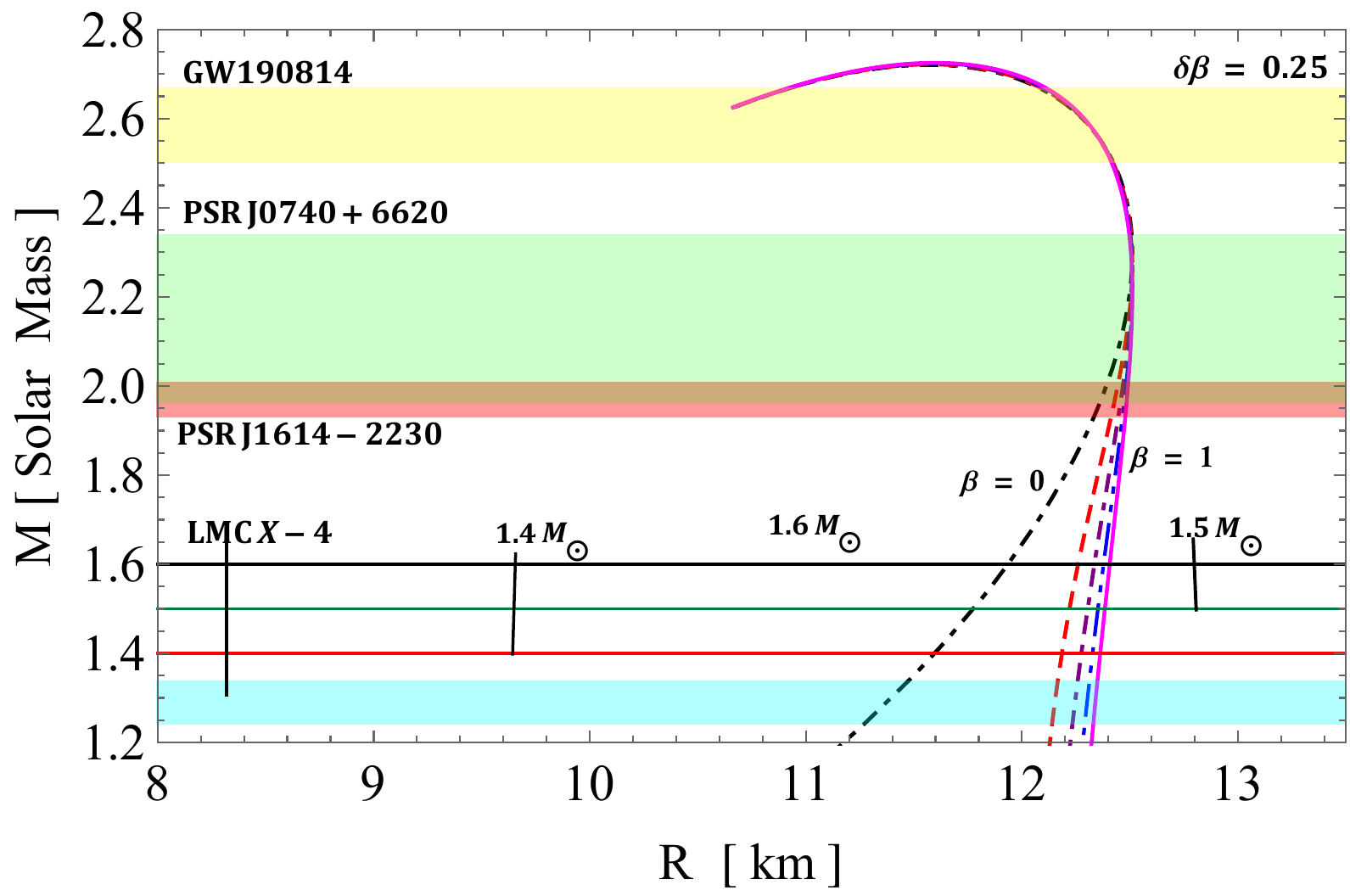}
    \caption{\textit{\textbf{Top Panels}:} The left and right panels show the $M-R$ curves depending on different values $\alpha_1$ and  $\beta$, respectively when $\theta^0_0=\rho$. \textit{\textbf{Right Panels}:} The left and right panels show the $M-R$ curves depending on different values of $\alpha_1$ and  $\beta$, respectively when $\theta^1_1=p_r$. } 
    \label{fig4}
\end{figure*}

\begin{figure*}[!htb]
    \centering
    \includegraphics[width=4.8cm,height=4cm]{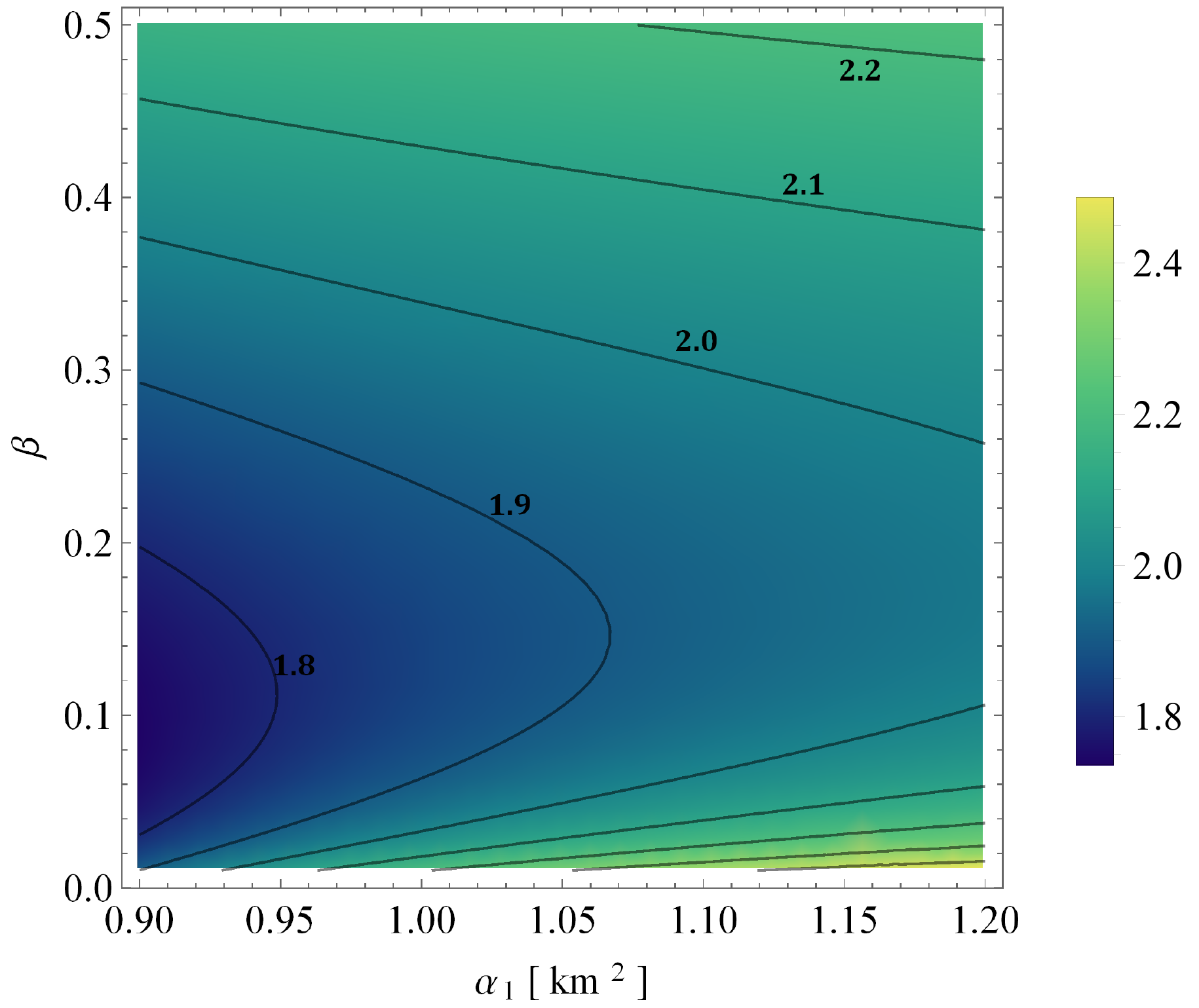}\includegraphics[width=5cm,height=4cm]{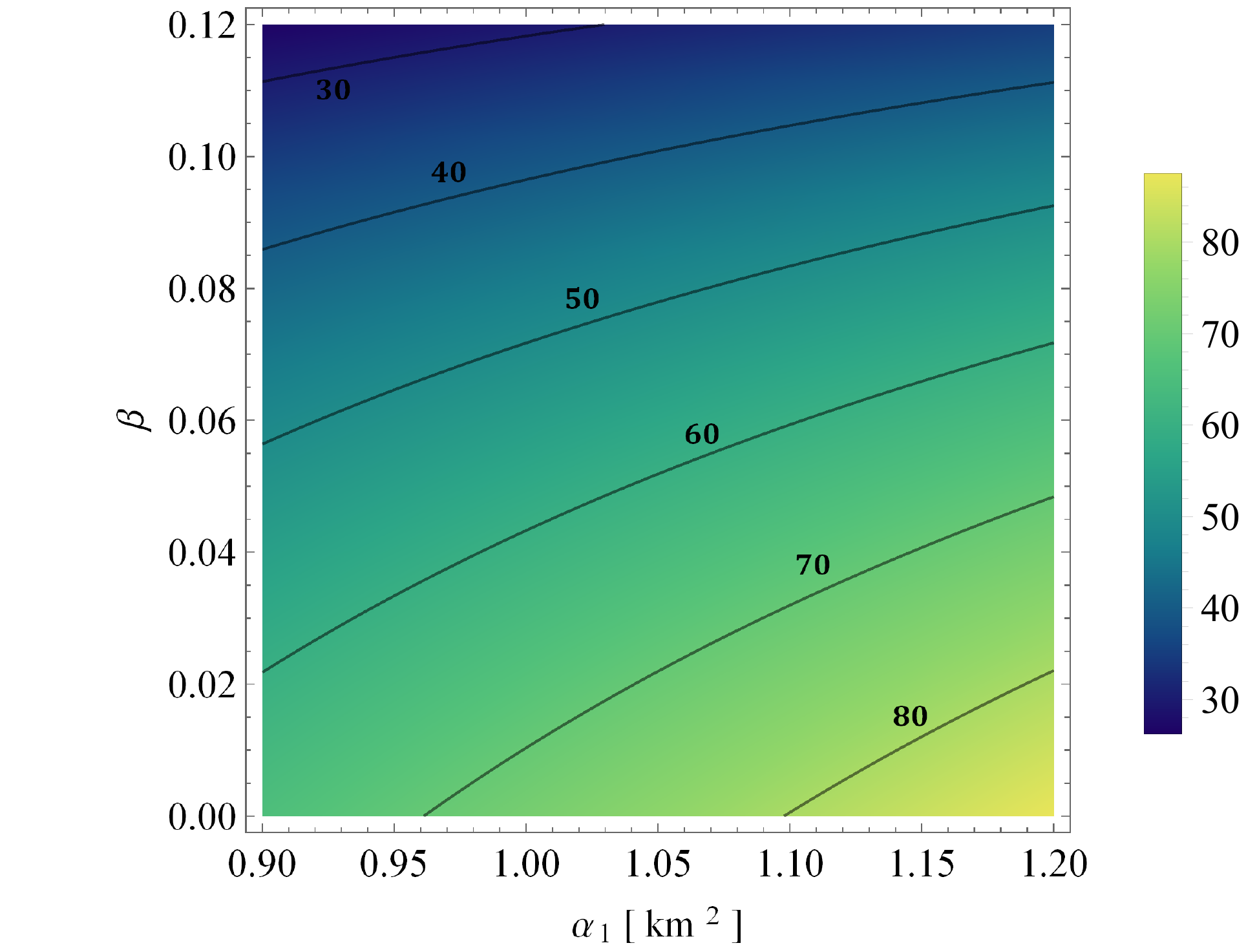}~
    \includegraphics[width=4.3cm,height=4cm]{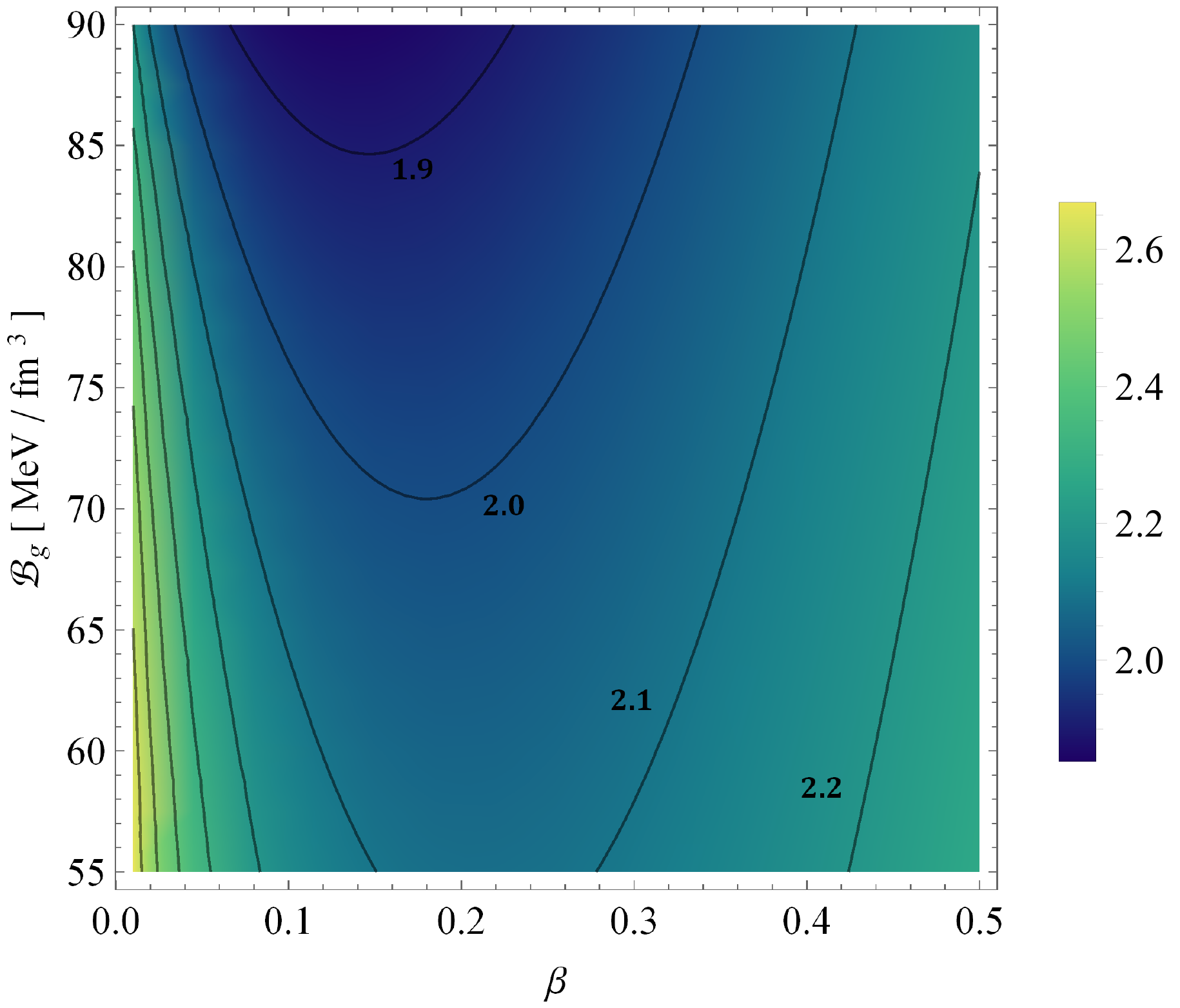}~
    \includegraphics[width=4.3cm,height=4cm]{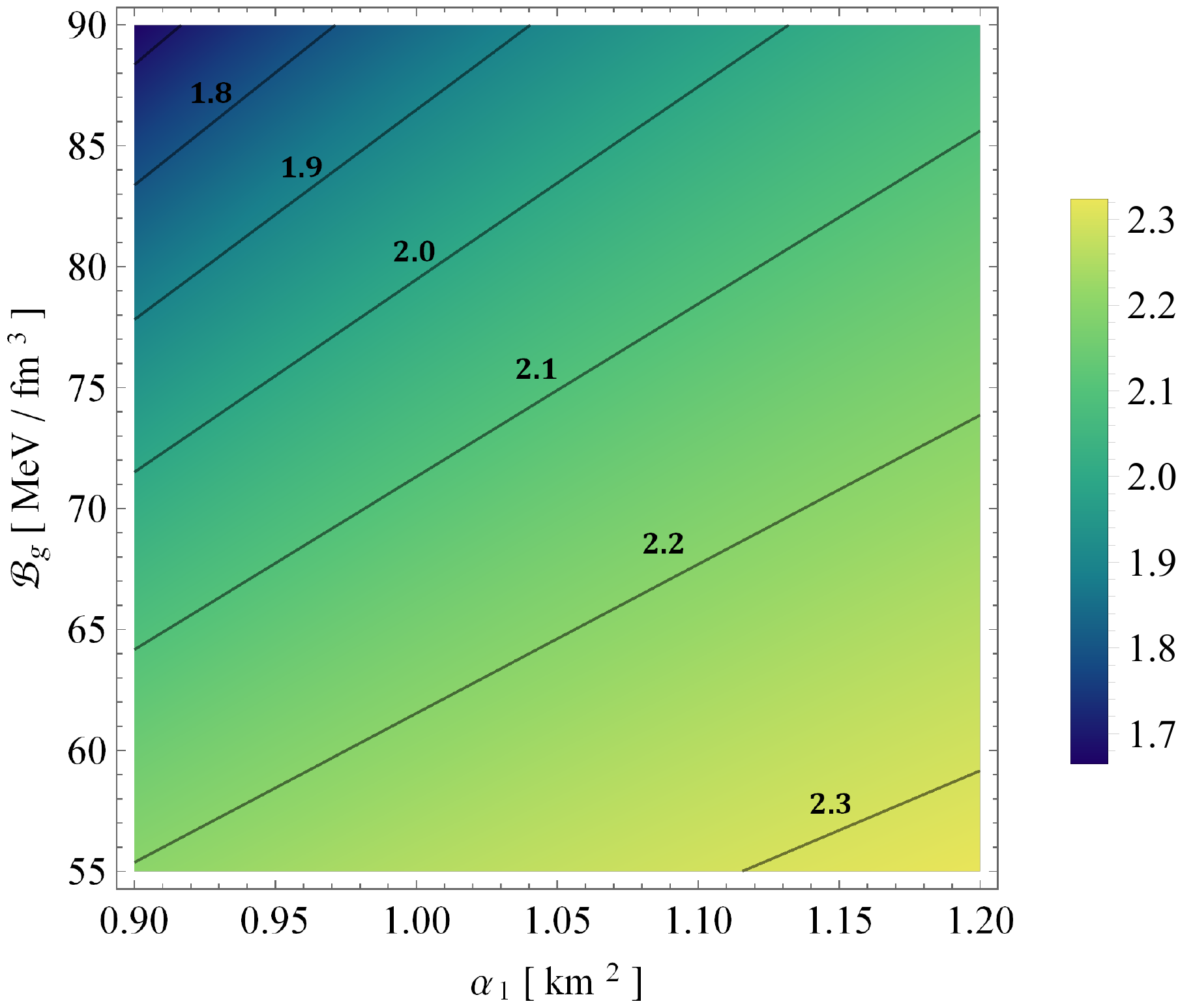}
    \caption{\textit{\textbf{For $\theta^0_0=\rho$ Solution}}:  $\alpha_1-\beta$ plane for equi-mass with $\mathcal{B}_g=65\,MeV/fm^3$ (\textit{first panel}) and $\alpha_1-\beta$ plane for equi-$\mathcal{B}_g$ (\textit{second panel}). The third panel shows $\mathcal{B}_g-\alpha_1$ plane for equi-mass for $\beta =0.06$, while \textit{fourth panel} is plotted for $\mathcal{B}_g-\beta$ plane  equi-mass with $\alpha_1 =1.05$. Where $R =11.5~km~ and ~N=5\times 10^{-5}/km^4$.}
    \label{fig5}
\end{figure*}

\begin{figure*}[!htb]
    \centering
    \includegraphics[width=4.5cm,height=4cm]{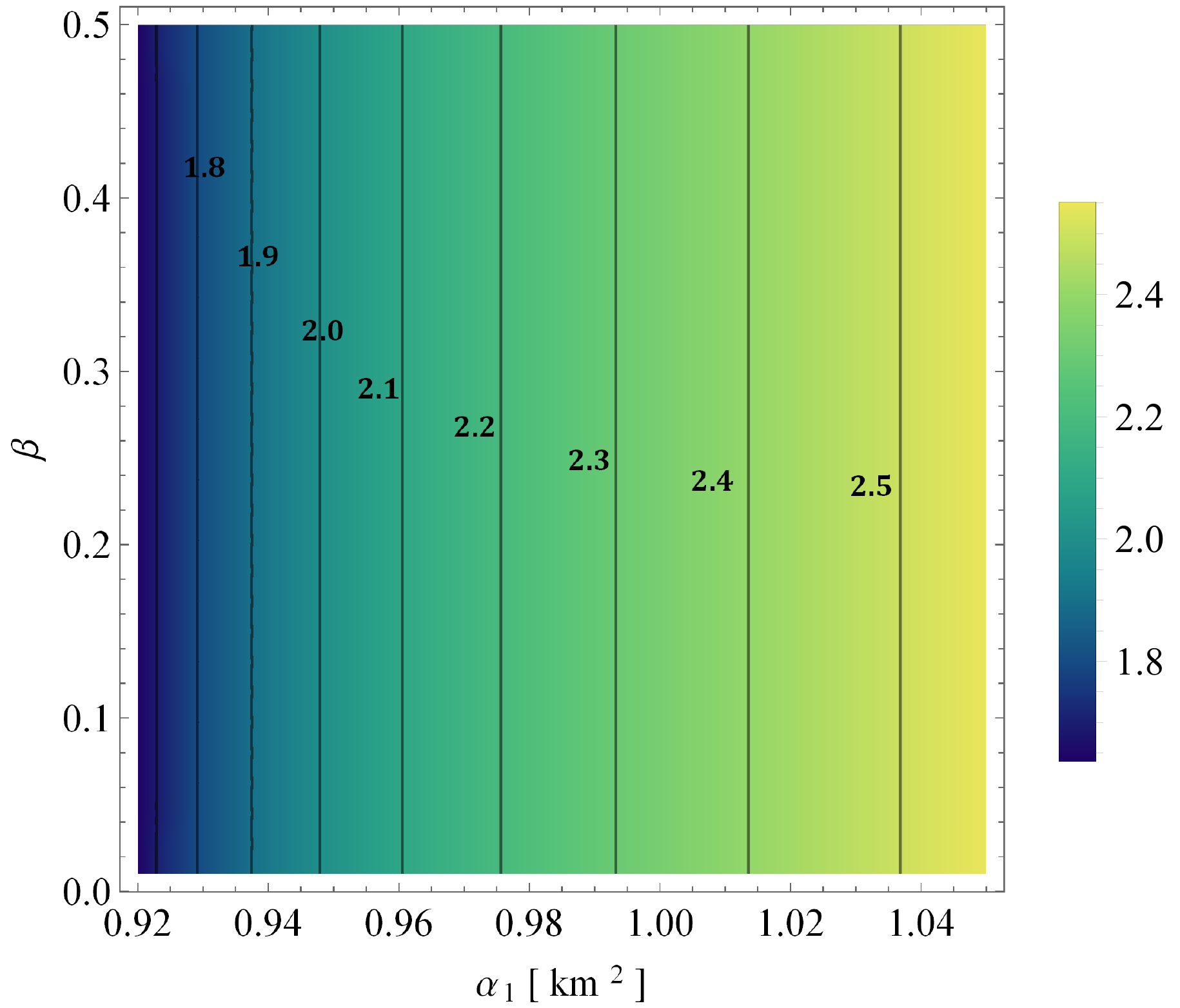}~
    \includegraphics[width=4.5cm,height=4cm]{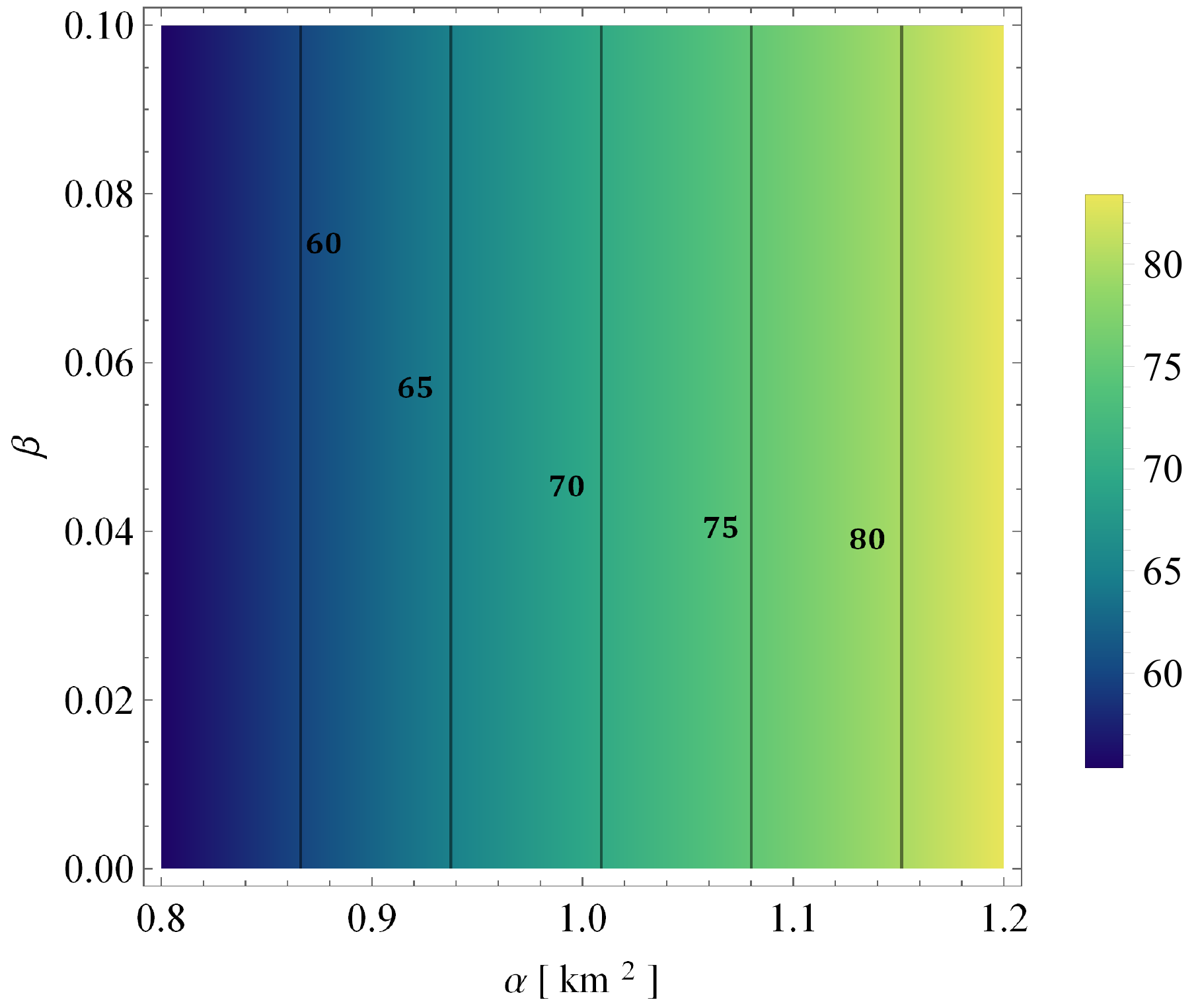}~
    \includegraphics[width=4.5cm,height=4cm]{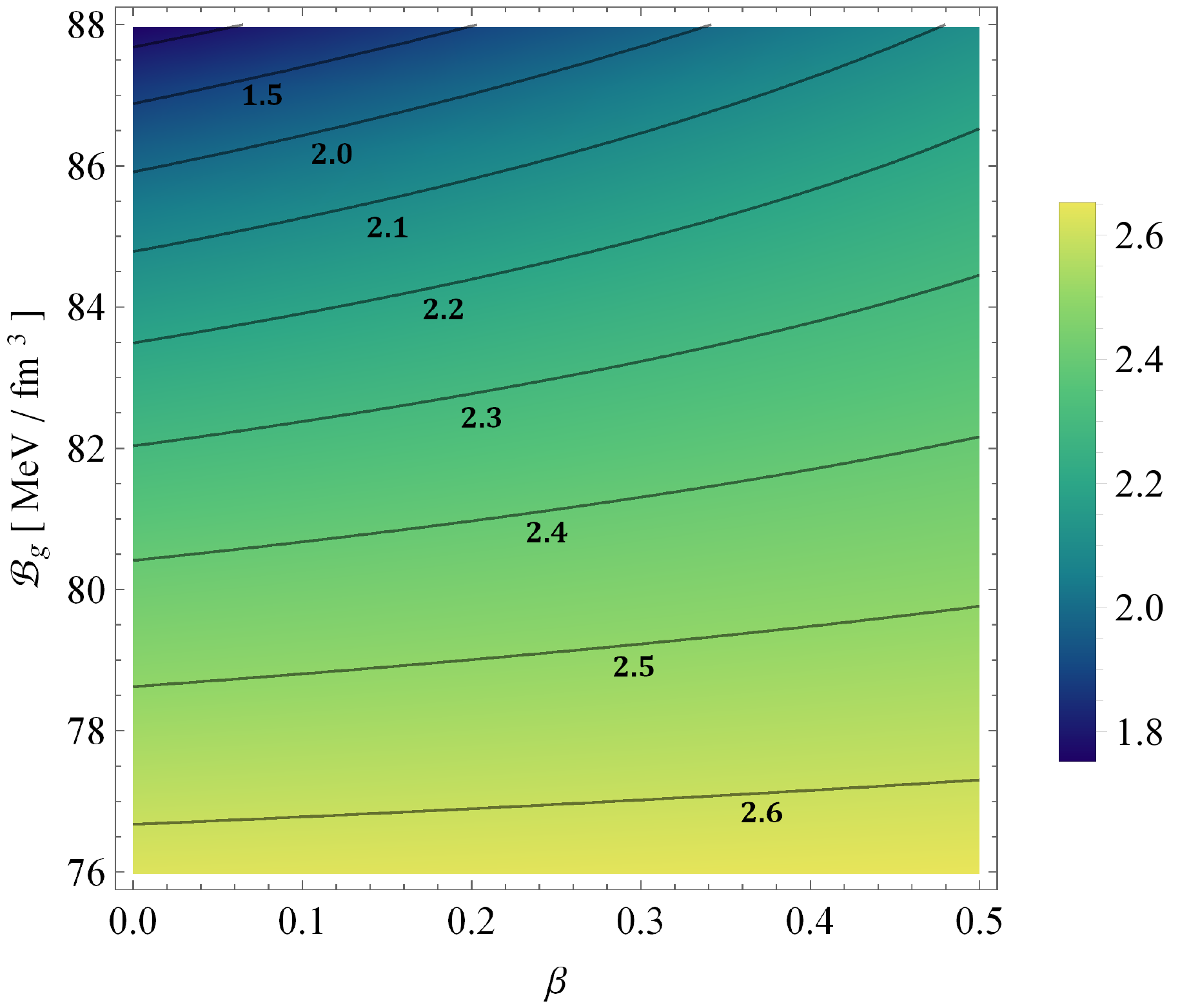}~
    \includegraphics[width=4.5cm,height=4cm]{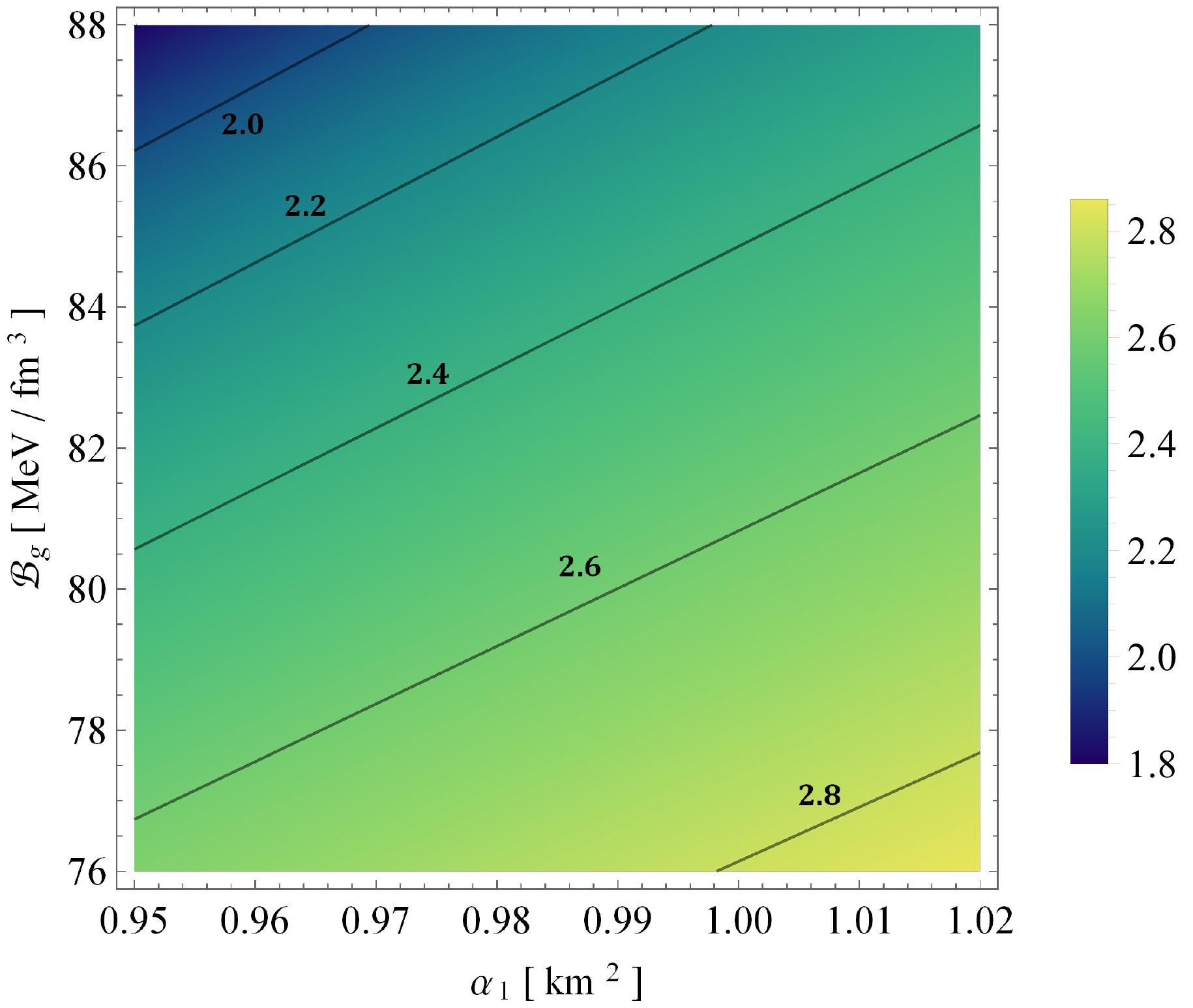}
    \caption{\textit{\textbf{For $\theta^1_1=p_r$ Solution}}:  $\alpha_1-\beta$ plane for equi-mass with $\mathcal{B}_g=65\,MeV/fm^3$ (\textit{first panel}) and $\alpha_1-\beta$ plane for equi-$\mathcal{B}_g$ (\textit{second panel}). The third panel shows $\mathcal{B}_g-\alpha_1$ plane for equi-mass for $\beta =0.06 $, while \textit{fourth panel} is plotted for $\mathcal{B}_g-\beta$ plane  equi-mass with $\alpha_1 =0.95 $. Where $R =11.5~km ~and~N=3\times 10^{-5}/km^4$.}
    \label{fig6}
\end{figure*}

The physical acceptability of the solution can be strengthened further by fitting the observational constraints from some recent gravitational wave sources like GW 170817. The GW 170817 event provides constraints on the mass and radius of neutron stars. For neutron stars of mass $1.4M_{\odot}$ should have $R>11.0^{+0.9}_{-0.6}$ km \cite{cap}, for $M=1.5M_{\odot}$ the radius must lie between ($11.8-13.1$) km and for $M=1.6 M_\odot$, $R>10.68^{+0.15}_{-0.03}$ km \cite{bau}.  These constraints are well fitted with the $M-R$ curves generated from the solution (see in Figure \ref{fig4}). Hence, we can use these $M-R$ curves to predict the radii of few well-known compact stars (see Tables \ref{table2} and \ref{table3}).
It can be seen that when both $f(Q)$ parameter $\alpha_1$ and the MGD coupling $\beta$ increase, the radii of the compact stars also increases. Also, the neutron star associated with the event GW 190817 predicted its mass very precisely that lies within $(2.5-2.67)M_\odot$, however, this event did not predict its radius. Hence, we use our $M-R$ curve to predict its possible radius. From the solution one i.e. $\theta_0^0=\rho$, $\alpha_1=1$ and $\beta=0.05$ gives a GR+CGD theory which predict its radius about $13.45_{-0.02}^{+0.01}\,km$ and it is strongly affected once the $f(Q)$ contribution is switched-on (i.e. $\alpha_1 \neq 1$ or $f(Q)$+CGD). Further, we can also see that in pure $f(Q)$ gravity i.e. $\alpha_1 \neq 1$ and $\beta =0$, its radius is about $12.58_{-0.16}^{+0.09}\, km$. As the CGD is turn-on, its radius started increasing. Hence, overall we can see that the $f(Q)$ parameter $\alpha_1$ may increase/decrease the radius however, the CGD parameter $\beta$ always increases the radius. This further means that the corresponding EoS can become softer when $\alpha_1<1$ and stiffer when $\alpha_1>1$ if compare with the GR case. However, the triggering of CGD only makes the EoS stiffer.


\begin{table*}[!htb]
\centering
\caption{The predicted radii of compact stars LMC X-4, PSR J1614-2230, PSR J0740+6620, GW190814 and GW 170817 for the case $\theta^0_0(r)=\rho(r)$ }\label{table2}
 \scalebox{0.67}{\begin{tabular}{| *{12}{c|} }
\hline
\multirow{3}{*}{Objects} & \multirow{3}{*}{$\frac{M}{M_\odot}$}   & \multicolumn{5}{c|}{{Predicted $R$ km}} & \multicolumn{5}{c|}{{Predicted $R$ km}} \\
\cline{3-12}
&& \multicolumn{5}{c|}{$\alpha_1$} & \multicolumn{5}{c|}{$\beta$} \\
\cline{3-12}
&  & 0.9 & 0.95 & 1.0 & 1.05 & 1.1 & 0 & 0.025 & 0.05 & 0.075 & 1   \\ \hline
LMC X-4 \cite{star1}  &  1.29 $\pm$ 0.05  & $11.38_{-0.11}^{+0.11}$  & $11.65_{-0.04}^{+0.05}$ & $11.84_{-0.12}^{+0.11}$ & $12.06_{-0.1}^{+0.1}$ & $12.28_{-0.05}^{+0.08}$ & $11.55_{0.06-}^{+0.08}$ & $11.56_{-0.1}^{+0.11}$ & $11.58_{-0.1}^{+0.1}$ & $11.64_{-0.13}^{+0.11}$ & $11.67_{-0.13}^{+0.1}$ \\
\hline
 PSR J1614-2230\cite{star2} & 1.97$\pm$0.04  & $12.48_{-0.05}^{+0.04}$ & $12.77_{-0.05}^{+0.04}$ & $13.03_{-0.03}^{+0.06}$ & $13.3_{-0.04}^{+0.06}$ & $13.55_{-0.05}^{+0.06}$ & $12.57_{-0.04}^{+0.02}$ & $12.66_{-0.04}^{+0.0.03}$ & $12.73_{-0.1}^{+0.03}$ & $12.75_{-0.04}^{+0.03}$ & $12.87_{-0.05}^{+0.04}$ \\
\hline
PSR J0740+6620 \cite{star3} & $2.14^{+0.2}_{-0.17}$ & $12.64_{-0.16}^{+0.1}$ & $12.93_{-0.17}^{+0.14}$ & $13.22_{-0.19}^{+0.16}$ & $13.49_{-0.02}^{+0.17}$ & $13.76_{-0.22}^{+0.19}$ & $12.69_{-0.12}^{+0.05}$ & $12.8_{-0.16}^{+0.08}$ & $12.88_{-0.16}^{+0.14}$ & $12.99_{-0.11}^{+0.14}$ & $13.07_{-0.21}^{+0.16}$ \\
\hline
GW190814 \cite{wenbin} & 2.5-2.67 & $12.74_{-0.07}^{+0.02}$ & $13.17_{-0.04}^{+0.01}$  & $13.45_{-0.02}^{+0.01}$ & $13.78_{-0.02}^{+0.01}$ & $14.09._{-0.02}^{+0.02}$ & $12.58_{-0.16}^{+0.09}$ & $12.84_{-0.08}^{+0.04}$ & $13.04_{-0.03}^{+0.01}$ & $13.2_{-0.01}^{+0.01}$ & 13.35$_{-0.03}^{+0.01}$ \\ \hline
GW 170817 & 1.4 & 11.62 & 11.86 & 12.09 & 12.36 & 12.53 & 11.77 & 11.8 & 11.81 & 11.87 & 11.91\\
\hline
GW 190814 & 1.5 & 11.81 & 12.05 & 12.29 & 12.53 & 12.74 & 11.95 & 12.0 & 12.02 & 12.07 & 13.12  \\
\hline
GW 170817 & 1.6 & 11.98 & 12.25 & 12.49 & 12.71 & 12.95 & 12.11 & 12.17 & 12.2 & 12.25 & 12.31\\
\hline
\end{tabular}}
\end{table*}
\begin{table*}[!htb]
\centering
\caption{The predicted radii of compact stars LMC X-4, PSR J1614-2230, PSR J0740+6620, GW190814 and GW 170817 for the case $\theta^1_1(r)=p_r(r)$ }\label{table3}
 \scalebox{0.67}{\begin{tabular}{|*{12}{c|} }
\hline
\multirow{3}{*}{Objects} & \multirow{3}{*}{$\frac{M}{M_\odot}$}   & \multicolumn{5}{c|}{{Predicted $R$ km}} & \multicolumn{5}{c|}{{Predicted $R$ km}} \\
\cline{3-12}
&& \multicolumn{5}{c|}{$\alpha_1$} & \multicolumn{5}{c|}{$\beta$} \\
\cline{3-12}
&  & 0.9 & 0.95 & 1.0 & 1.05 & 1.1 & 0 & 0.025 & 0.05 & 0.075 & 1   \\ \hline
LMC X-4 \cite{star1}  &  1.29 $\pm$ 0.05  & $11.67_{-0.02}^{+0.01}$  & $12.02_{-0.01}^{+0.01}$ & $12.37_{-0.01}^{+0.01}$ & $12.7_{-0.01}^{+0.01}$ & $13.04_{-0.01}^{+0.01}$ & $11.36_{0.09-}^{+0.11}$ & $12.15_{-0.01}^{+0.02}$ & $12.25_{-0.02}^{+0.01}$ & $12.30_{-0.01}^{+0.01}$ & $12.34_{-0.01}^{+0.01}$ \\
\hline
 PSR J1614-2230\cite{star2} & 1.97$\pm$0.04  & $12.10_{-0.01}^{+0.01}$ & $12.41_{-0.03}^{+0.02}$ & $12.71_{-0.02}^{+0.02}$ & $12.98_{-0.03}^{+0.03}$ & $13.25_{-0.02}^{+0.03}$ & $12.37_{-0.03}^{+0.02}$ & $12.43_{-0.02}^{+0.0.02}$ & $12.46_{-0.01}^{+0.01}$ & $12.47_{-0.01}^{+0.01}$ & $12.49_{-0.01}^{+0.01}$ \\
\hline
PSR J0740+6620 \cite{star3} & $2.14^{+0.2}_{-0.17}$ & $12.19_{-0.03}^{+0.01}$ & $12.50_{-0.09}^{+0.03}$ & $12.80_{-0.01}^{+0.06}$ & $13.08_{-0.11}^{+0.08}$ & $13.36_{-0.09}^{+0.1}$ & $12.47_{-0.10}^{+0.03}$ & $12.50_{-0.07}^{+0.01}$ & $12.50_{-0.05}^{+0.01}$ & $12.50_{-0.03}^{+0.1}$ & $12.50_{-0.02}^{+0.01}$ \\
\hline
GW190814 \cite{wenbin} & 2.5-2.67 & $11.77_{-0.001}^{+0.23}$ & $12.31_{-0.24}^{+0.12}$  & $12.74_{-0.14}^{+0.41}$ & $13.11_{-0.08}^{+0.04}$ & $13.47_{-0.06}^{+0.01}$ & $12.31_{-0.21}^{+0.11}$ & $12.31_{-0.21}^{+0.11}$ & $12.31_{-0.21}^{+0.11}$ & $12.31_{-0.21}^{+0.11}$ & $12.31_{-0.21}^{+0.11}$ \\ 
\hline
GW 170817 & 1.4 & 11.71 & 12.04 & 12.39 & 12.71 & 13.04 & 11.59 & 12.18 & 12.28 & 12.33 & 12.36\\
\hline
GW 190814 & 1.5 & 11.76 & 12.09 & 12.42 & 12.74 & 13.06 & 11.78 & 12.22 & 12.31 & 12.35 & 12.39  \\
\hline
GW 170817 & 1.6 & 11.83 & 12.15 & 12.47 & 12.77 & 13.09 & 11.93 & 12.26 & 12.34 & 12.38 & 12.41\\
\hline
\end{tabular}}
\end{table*}
\subsection{Measurements of the constraints on total mass and Bag constant value via equi-plane diagrams} \label{sec5.5} 
In this section, we discuss the analysis of Figures \ref{fig5} and \ref{fig6}. 
The Figure \ref{fig5} is plotted for the solution \ref{solA} i.e. ($\rho=\theta^0_0$) by taking $R=11.5\,km$. The first panel of this Figure \ref{fig5} describes the equi-mass contours in the $\alpha_1-\beta$ plane which is plotted by taking $\mathcal{B}_g=65\,MeV/fm^3$.  From this figure, we see that when $\alpha_1$ fixed, then we have two scenarios: (i) if $\alpha_1$ is low i.e. $\alpha_1=0.9$, then there is no significant change in mass for $0\le \beta\le 0.3$. But when $\beta>0.3$, then mass start increasing slightly but it remains lower than $2.15 M_{\odot}$. (ii) If $\alpha_1$ is high i.e. $\alpha_1=1.2$, then maximum mass is achieved for lower values of $\beta$ which is more than $2.4 M_{\odot}$ but when start increasing $\beta$ up $\sim$ 0.3, then mass decreases up to $2 M_{\odot}$, and beyond this value of $\beta$, the mass again start increasing and reach up to $\simeq$ $2.2 M_{\odot}$ for $\beta=0.5$. This shows that the maximum mass can be achieved on $\alpha_1-\beta$ plane for the high $\alpha_1$ and low $\beta$. The second panel shows the $\alpha_1-\beta$ plane for equi-Bag constant contour diagram  with $\mathcal{B}_g=65\,MeV/fm^3$. We can observe that the Bag constant $\mathcal{B}_g$ takes lower value for higher $\beta$ with any $\alpha_1$ but it attains maximum at higher $\alpha_1$ and lower value of $\beta$ i.e. $(\alpha_1,\beta)\cong (1.2, 0.0) $, and maximum value of $\mathcal{B}_g$ is $\simeq$ $86 MeV/fm^3$. Now when we look at the third panel, which is shown for the equi-mass contour diagram on $\beta-\mathcal{B}_g$ plane with fix $\alpha_1=1.05$. It can be seen that when we increase $\beta$ for fix $\mathcal{B}_g$, the mass decreases up to $\beta\simeq 0.35$ and start increasing slightly beyond this value while there is no significant change in mass when we vary $\mathcal{B}_g$ with all $\beta \ge 0.12$. This implies that the model will take a higher mass for the lower value of $\mathcal{B}_g$ and $\beta$. Finally the last right panel shows that behavior of equi-mass on the $\alpha_1-\mathcal{B}_g$ plane for $\beta=0.06$. This figure indicates that when we fix $\alpha_1$  and increase $\mathcal{B}_g$, then the mass is decreasing while it increases for increasing $\alpha_1$ with fix $\mathcal{B}_g$. The maximum mass value $(2.3M_{\odot})$ is achieved at ($\alpha_1, \mathcal{B}_g) \cong (1.2, 55)$. \\

Now we move on the Figure \ref{fig6} which is plotted for the solution \ref{solB} i.e. ($p_r=\theta^1_1$) by taking $R=11.5\,km$. As we can see that the first panel represents the equi-mass on the $\alpha_1-\beta$ plane with $\mathcal{B}_g=65\,MeV/fm^3$. It can be clearly observed that mass is independent of $\beta$ for any fix value of $\alpha_1$, while it is totally dependent on $\alpha_1$ i.e. when we increase $\alpha_1$ for any fix value of $\beta$, then mass also increases and maximum value of mass is obtained $M\simeq 2.55M_{\odot}$ at $\alpha_1\simeq 1.05$. Furthermore, the contour diagram for equi-$\mathcal{B}_g$ on the $\alpha_1-\beta$ plane is shown in the second panel of this figure. We can observe that the Bag constant $\mathcal{B}_g$ shows the same behavior as mass i.e. the $\mathcal{B}_g$ is dependent only on the parameter $\alpha_1$. The Bag constant value is increasing from $55~ MeV/fm^3$ to $85~ MeV/fm^3$ when $\alpha_1$ increases $0.8$ to $1.2$. Furthermore,  if we look at the third panel for the equi-mass contour diagram in  $\beta-\mathcal{B}_g$ plane for fix $\alpha_1=0.95$, we observed that the mass is independent of decoupling constant $\beta$ while it is dependent on $\mathcal{B}_g$. The mass is decreasing when $\mathcal{B}_g$ increases and the maximum mass value $2.65M_{\odot}$ is achieved when $\mathcal{B}_g=76 MeV/fm^3$. The last panel of this Figure \ref{fig6} is plotted for equi-mass contour diagram in  $\alpha_1-\mathcal{B}_g$ plane with fix $\beta=0.06$ which shows that mass is dependent on both parameters $\alpha_1$ and $\mathcal{B}_g$. The maximum mass value $\approx 2.85M_{\odot}$ is obtained at higher value of $\alpha_1$ with lower $\mathcal{B}_g$ i.e. $(\alpha_1, \mathcal{B}_g)\equiv (1.02, 76)$.

\section{Concluding Remarks} \label{sec6}
In the present work, we have modeled the astrophysical compact star objects along with recent observations of GW190814 and GW 170817 events. Our main objective of this work is to model the compact objects with mass beyond the $2M_{\odot}$ in the framework of $f(Q)$ theory of gravity as well as to explore the effects of anisotropy introduced via CGD. To solve the system of equations we apply the Tolman IV ansatz corresponding to the temporal component of the metric function together with the stage star equation of state (EoS), in particular, MIT Bag EoS.  The field equations due to gravitational decoupling have been solved by  taking two approaches, known as the mimic constraints approach, (i) $\theta^0_0=\rho$ - sector and (ii) $\theta^1_1=p_r$ - sector, which are mentioned in sections \ref{solA} and \ref{solB}. The physical viability of the obtained  models have been tested under the nonmetricity parameter, $\alpha_1$ and the decoupling constant, $\beta$. 

In our analysis, we obtained that the parameters $\alpha_1$ and $\beta$ show some influence on the energy density and pressure profiles of the model. It is observed that the nonmetricity parameter $\alpha_1$ increases the densities and stresses within the stellar configurations. On the other hand, the decoupling parameter also enhances the density of the compact objects, particularly at the stellar surfaces for both solutions. 
The anisotropy $\Delta^{\text{eff}}$ is positive and increasing throughout the star which introduces a repulsive force. This repulsive force stabilizes the object against a gravitational force acting in the inward direction.  Moreover, we also noted the maximum amount of anisotropy in magnitude generated in the $\theta^0_0=\rho$ models is $70\%$ greater than the $\theta^1_1=p_r$ counterparts. 

The mass profiles for both solutions are also physically well-behaved. The originality of our current work can be observed from $M-R$ curves shown in Figure \ref{fig5}. In these $M-R$ curves, we have fitted the masses of well-known compact objects such as LMC X-4, PSR J1614-2230, and PSR J0740+6620 and predicted corresponding radii under both solutions. Furthermore, we also fitted the objects with masses above than 2.0 $M_{\odot}$ and predicted their radii.  
For the $\theta^0_0=\rho$ solution, predicted radii for the above compact objects are lying between $11.3-14.1$ km while in the $\theta^1_1=p_r$ models, radii for similar compact objects ranging between $11.5-13.5$ km. In particular, the predicted radii for the objects with 2.0 $M_{\odot}$ is $12.48_{-0.05}^{+0.04}$ km and $13.55_{-0.05}^{+0.06}$ km when $\alpha_1=0.9$ and $\alpha_1=1.1$, respectively for $\theta^0_0(r)=\rho(r)$ case while $12.10_{-0.01}^{+0.01}$ km and $13.25_{-0.02}^{+0.03}$ km for $\theta^1_1(r)=p_r(r)$ case. 
On the other hand, the contributions from the metricity factor in both solutions, nonmetricity $Q$ allows for bigger self-gravitating compact objects (larger radii) with larger masses. However, if we compare the GW190814 event, the effect of changing the decoupling constant with fixed $\alpha_1$ yields slightly lower radii as compared to varying the nonmetricity parameter $\alpha_1$ with fixed $\beta$. 
Based on the above analysis, we observe that a combination of contributions from the nonmetricity factor and the decoupling parameter allows for higher mass neutron stars that are stable and may be able to describe the secondary ancestor of the black hole-neutron star observed in the GW190814 event.

\section*{Data Availability}
The data underlying this article are available in the article and in its online supplementary material.

\begin{widetext}
\section*{Appendix}
\begin{small}
\begin{eqnarray}
&&\hspace{0.0cm} f(r)=\big(L r^2+N r^4+1\big)^2,~~ \theta_{10}(r)=2 \alpha_2 N^2 r^7-3 \alpha_1 N^2 r^5+2 \alpha_2 N r^3-17 \alpha_1 N r,\nonumber\\
&&\hspace{0.0cm}\theta_{30}(r)=L^2 \big(\alpha_2 r^3-\alpha_1 r\big)+L r \big(\alpha_2+3 \alpha_2 N r^4-4 \alpha_1 N r^2\big),~~
\theta_{12}(r)=2 L \big(-3 \alpha_1+\alpha_2 N r^6-2 \alpha_1 N r^4+\alpha_2 r^2\big),\nonumber\\
&&\hspace{0.0cm} p_{t1}(r)={-4 {Bg} f(r) \big(\alpha_1 \big(4 L^2 r^4+L r^2 \big(8 N r^4+15\big)+4 N^2 r^8+20 N r^4+6\big)-2 {\alpha_2} r^2\,f(r) \big)}+8 {Bg}^2 r^2 f^2(r)+12 {\alpha_2}^2 N^2 r^{10}\nonumber\\&&\hspace{0.5cm} +L^3 r^4 \big(4 {\alpha_1}^2 (8 N r^4+9)+8 {\alpha_2}^2 r^4 \big(N r^4+1\big)-{\alpha_1} {\alpha_2} r^2 \big(32 N r^4+37\big)\big) +44 {\alpha_1}^2 N^2 r^6+L^2 r^2 \big[(2 {\alpha_1}^2 (24 N^2 r^8+64 N r^4+15)\nonumber\\&&\hspace{0.5cm} -{\alpha_1} {\alpha_2} r^2(48 N^2 r^8+121 N r^4+53)+12 r^4 \big({\alpha_2}+{\alpha_2} N r^4\big)^2\big] +L \Big(2 {\alpha_1}^2 \big(16 N^3 r^{12}+74 N^2 r^8+45 N r^4+9\big)-{\alpha_1} {\alpha_2} r^2 \big(32 N^3 r^{12} \nonumber\\&&\hspace{0.5cm} +131 N^2 r^8+126 N r^4+27\big)+8 {\alpha_2}^2 r^4 (N r^4+1)^3\Big)-8 {\alpha_1} {\alpha_2} N^4 r^{16}+8 {\alpha_1}^2 N^4 r^{14}+8 {\alpha_2}^2 N^3 r^{14}-47 {\alpha_1} {\alpha_2} N^3 r^{12}+56 {\alpha_1}^2 N^3 r^{10}\nonumber\\&&\hspace{0.5cm}+2 L^4 r^6 \big({\alpha_2} r^2-2 {\alpha_1}\big)^2\Big],\nonumber
\end{eqnarray}
\begin{eqnarray}
&&\hspace{0.0cm} \theta_{11}(r)=\Big[2 L^3 \big(\alpha_2^2 r^8-8 \alpha_1 \alpha_2 r^6+12 \alpha_1^2 r^4\big)+L^2 r^2 \big(18 \alpha_1^2 \big(2 \beta +4 N r^4+3\big)+6 \alpha_2^2 r^4 \big(N r^4+1\big)-\alpha_1 \alpha_2 r^2 \big(6 \beta +48 N r^4+37\big)\big)\nonumber\\&&\hspace{0.5cm}+2 L \big(\alpha_1^2 \big(36 N^2 r^8+6 (9 \beta +10) N r^4-9\big)-\alpha_1 \alpha_2 r^2 \big(3 (\beta +4)+24 N^2 r^8+(9 \beta +38) N r^4\big)+3 r^4 \big(\alpha_2+\alpha_2 N r^4\big)^2\big)+2 \alpha_2^2 N^3 r^{14}\nonumber\\&&\hspace{0.5cm}-16 \alpha_1 \alpha_2 N^3 r^{12}+24 \alpha_1^2 N^3 r^{10}+6 \alpha_2^2 N^2 r^{10}-12 \alpha_1 \alpha_2 \beta  N^2 r^8-39 \alpha_1 \alpha_2 N^2 r^8+72 \alpha_1^2 \beta  N^2 r^6+66 \alpha_1^2 N^2 r^6+6 \alpha_2^2 N r^6\nonumber\\&&\hspace{0.5cm}-12 \alpha_1 \alpha_2 \beta  N r^4-26 \alpha_1 \alpha_2 N r^4-54 \alpha_1^2 N r^2+2 \alpha_2^2 r^2\Big],\nonumber\\
&&\hspace{0.0cm}  \theta_{20}(r)=\Big[\big(4 \mathcal{B}_g r^2\,f(r) -\alpha_1 \big(4 L^2 r^4+L r^2 \big(6 \beta +8 N r^4+9\big)+4 N^2 r^8+(12 \beta +11) N r^4+3\big)+2 \alpha_2 r^2\,f(r)\big)-2 r^2 \big(L+2 N r^2\big) \nonumber\\&&\hspace{0.5cm} \times \big(\alpha_2+8 \mathcal{B}_g \,f(r)+L^2 \big(\alpha_2 r^4-2 \alpha_1 r^2\big)+2 L \big(-9 \alpha_1+\alpha_2 N r^6-2 \alpha_1 N r^4+\alpha_2 r^2\big)+\alpha_2 N^2 r^8-2 \alpha_1 N^2 r^6+2 \alpha_2 N r^4-34 \alpha_1 N r^2\big) \nonumber\\&&\hspace{0.5cm}\times \big(4 \mathcal{B}_g\, r^2\,f(r)-\alpha_1 \big(4 L^2 r^4+L r^2 \big(6 \beta +8 N r^4+9\big)+4 N^2 r^8+(12 \beta +11) N r^4+3\big)+2 \alpha_2 \big(L r^3+N r^5+r\big)^2\big)-r \big(L r^2+N r^4+1\big) \nonumber\\&&\hspace{0.5cm}\times \big(\alpha_2+8 \mathcal{B}_g \big(L r^2+N r^4+1\big)^2+L^2 \big(\alpha_2 r^4-2 \alpha_1 r^2\big)+2 L \big(-9 \alpha_1+\alpha_2 N r^6-2 \alpha_1 N r^4+\alpha_2 r^2\big)+\alpha_2 N^2 r^8-2 \alpha_1 N^2 r^6+2 \alpha_2 N r^4\nonumber\\&&\hspace{0.5cm}-34 \alpha_1 N r^2\big)  \times \Big[8 \mathcal{B}_g \big(3 L r^2+5 N r^4+1\big) \big(L r^3+N r^5+r\big)-2 \alpha_1 \big(8 L^2 r^3+3 L r \big(2 \beta +8 N r^4+3\big)+2 N r^3 \big(12 \beta +8 N r^4+11\big)\big)\nonumber\\&&\hspace{0.5cm}+4 \alpha_2 \big(3 L r^2+5 N r^4+1\big)  \big(L r^3+N r^5+r\big)\Big] +3 \big(L r^2+N r^4+1\big) \big(\alpha_2+8 \mathcal{B}_g\,f(r)+L^2 \big(\alpha_2 r^4-2 \alpha_1 r^2\big)+2 L \big(-9 \alpha_1+\alpha_2 N r^6\nonumber\\&&\hspace{0.5cm} -2 \alpha_1 N r^4+\alpha_2 r^2\big)+\alpha_2 N^2 r^8 -2 \alpha_1 N^2 r^6+2 \alpha_2 N r^4-34 \alpha_1 N r^2\big) \Big(4 \mathcal{B}_g\, r^2\,f(r)-\alpha_1 \big(4 L^2 r^4+L r^2 \big(6 \beta +8 N r^4+9\big)+4 N^2 r^8\nonumber\\&&\hspace{0.5cm}+(12 \beta +11) N r^4+3\big)+2 \alpha_2\, r^2\,f(r)\Big)\Big],\nonumber
\end{eqnarray}
\end{small}
\end{widetext}


\begin{thebibliography}{99}
\bibitem{Perlmutter99} S. Perlmutter, et al., \textit{The Astro. J} \textbf{1999}, 517, 565.

\bibitem{Riess98} A. G. Riess, et al., \textit{The Astro. J} \textbf{1998}, 116, 1009.

\bibitem{Benn82} I. M. Benn, T. Dereli, R. W. J. Tucker, \textit{J. Phys. A} \textbf{1982} 15, 849.

\bibitem{Hehl95} F. W. Hehl, et al., \textit{Phys. Repts.} \textbf{1995}, 258, 1.

\bibitem{Boulanger06} N. Boulanger, I. Kirsch, \textit{Phys. Rev. D} \textbf{2006}, 73, 124023.

\bibitem{Kirsch05} I. Kirsch, \textit{Phys. Rev. D} \textbf{2005}, 72, 024001.

\bibitem{Aldrovandi13} R. Aldrovandi, J. G. Pereira, Teleparallel Gravity, Vol. 173 (\textit{Springer, Dordrecht,} 2013)

\bibitem{Jimenez18} J. B. Jimenez, L. Heisenberg, T. Koivisto, \textit{Phys. Rev. D} \textbf{2018}, 98, 044048.

\bibitem{Conroy18} A. Conroy, T. Koivisto, \textit{Eur. Phys. J. C} \textbf{2018}, 78, 923.

\bibitem{Golovnev17} A. Golovnev, T. Koivisto, M. Sandstad, \textit{Class. Quantum Grav.} \textbf{2017} 34, 145013.

\bibitem{Jimenez16} J. B. Jimenez, T. S. Koivisto, \textit{Phys. Lett. B} \textbf{2016}, 756, 400.

\bibitem{Nester99} J. M. Nester, H. J. Yo, \textit{Chin. J. Phys.} \textbf{1999}, 37, 113.

\bibitem{Adak18} M. Adak, \textit{Int. J. Geom. Meth. Mod. Phys.} \textbf{2018}, 15, 1850198.

\bibitem{Weyl18} H. Weyl, \textit{S. Preuss. Akad. Wiss.} \textbf{1918}, 465, 1.

\bibitem{Einstein18} A. Einstein, \textit{S. Preuss. Akad. Wiss.} \textbf{1918}, 478, 1.

\bibitem{Cartan23} E. Cartan, \textit{Ann. Ec. Norm.} \textbf{1923} 40, 325; \textbf{1924} 41, 1; \textbf{1925} 42, 17.

\bibitem{Weitzenbock23} R. Weitzenb$\ddot{o}$ck, Invariantentheorie, noordhoff, groningen \textbf{1923}.

\bibitem{Bahamonde21} S. Bahamonde, K. F. Dialektopoulos, et al., \textit{arXiv:2106.13793} \textbf{2021}.

\bibitem{Cai16} Y. F. Cai, S. Capozziello, M. De Laurentis, E. N. Saridakis, \textit{Rept. Prog. Phys.} \textbf{2016}, 79, 106901.

\bibitem{Heisenberg19} L. Heisenberg, \textit{Phys. Rept} \textbf{2019}, 796, 1.

\bibitem{Jimenez20} J. B. Jimenez, et al., \textit{Phys. Rev. D} \textbf{2020}, 101, 103507.

\bibitem{Lu19} J. Lu, X. Zhao, G. Chee, \textit{Eur. Phys. J. C} \textbf{2019}, 79, 530.

\bibitem{Anagnostopoulos21} F. K. Anagnostopoulos, S. Basilakos, E. N. Saridakis, \textit{Phys. Lett. B} \textbf{2021}, 822, 136634.

\bibitem{Lazkoz19} R. Lazkoz, et al., \textit{Phys. Rev. D} \textbf{2019}, 100, 104027.

\bibitem{Lin21} R. H. Lin, X. H. Zhai, \textit{Phys. Rev. D} \textbf{2021}, 103, 124001.

\bibitem{Ambrosio22} F. D'Ambrosio, et al., \textit{Phys. Rev. D} \textbf{2022}, 105, 024042.

\bibitem{Wang22} W. Wang, H. Chen, T. Katsuragawa, \textit{Phys. Rev. D} \textbf{2022}, 105, 024060.

\bibitem{Latorre18} A. D. Latorre, G. J. Olmo, M. Ronco, \textit{Phys. Lett. B} \textbf{2018}, 780, 294.

\bibitem{Dialektopoulos19} K. F. Dialektopoulos, T. S. Koivisto, S. Capozziello, \textit{Eur. Phys. J. C} \textbf{2019}, 79, 606.

\bibitem{Barros20} B. J. Barros, et al., \textit{Phys. Dark Universe} \textbf{2020}, 30, 100616.

\bibitem{Bajardi20} F. Bajardi, D. Vernieri, S. Capozziello, \textit{Eur. Phys. J. Plus} \textbf{2020}, 135, 912.

\bibitem{Ambrosio20} F. D'Ambrosio, M. Garg, L. Heisenberg, \textit{Phys. Lett. B} \textbf{2020}, 811, 135970.

\bibitem{Ayuso21} I. Ayuso, R. Lazkoz, V. Salzano, \textit{Phys. Rev. D} \textbf{2021}, 103, 063505.

\bibitem{Flathmann21} K. Flathmann, M. Hohmann, \textit{Phys. Rev. D} \textbf{2021}, 103, 044030.

\bibitem{Fruscinate21} N. Frusciante, \textit{Phys. Rev. D} \textbf{2021}, 103, 044021.

\bibitem{Khyllep21} W. Khyllep, A. Paliathanasis, J. Dutta, \textit{Phys. Rev. D} \textbf{2021}, 103, 103521.

\bibitem{Capozziello:plb} S. Capozziello, R. D'Agostino, Phys. Lett. B 832 (2022) 137229 

\bibitem{Albuquerque22} I. S. Albuquerque, N. Frusciante, \textit{Phys. Dark Universe} \textbf{2022}, 35, 100980.

\bibitem{hans3} B. Chilambwe, S. Hansraj, S. D. Maharaj, \textit{Int. J. Modern Phys. D} \textbf{2015}, 24, 1550051.

\bibitem{hans1} S. Hansraj, B. Chilambwe, S. D. Maharaj, \textit{Eur. Phys. J. C} \textbf{2015}, 75, 277.

\bibitem{hans4} S. Hansraj, S. D. Maharaj, B. Chilambwe, \textit{Phys. Rev. D} \textbf{2019}, 100, 124029.

\bibitem{hans2} S. D. Maharaj, B. Chilambwe, S. Hansraj, \textit{Phys. Rev. D} \textbf{2015}, 91, 084049.

\bibitem{hans6} S. D. Maharaj, S. Hansraj, P. Sahoo, \textit{Eur. Phys. J. C} \textbf{2021}, 81, 1113.

\bibitem{OvallePRD2017} J. Ovalle, \textit{Phys. Rev. D} \textbf{2017}, 95, 104019.

\bibitem{OvallePLB2019} J. Ovalle, \textit{Phys. Lett. B} \textbf{2019}, 788, 213.

\bibitem{Contreras21}  \'A.~Rinc\'on, L.~Gabbanelli, E.~Contreras and F.~Tello-Ortiz,
%``Minimal geometric deformation in a Reissner\textendash{}Nordstr\"om background,''
Eur. Phys. J. C \textbf{79} (2019)  873

\bibitem{Ovalle:2018vmg}
J.~Ovalle and A.~Sotomayor,
%``A simple method to generate exact physically acceptable anisotropic solutions in general relativity,''
Eur. Phys. J. Plus \textbf{133} (2018)  428

\bibitem{Gabbanelli:2019txr}
L.~Gabbanelli, J.~Ovalle, A.~Sotomayor, Z.~Stuchlik and R.~Casadio,
%``A causal Schwarzschild-de Sitter interior solution by gravitational decoupling,''
Eur. Phys. J. C \textbf{79} (2019) 486

\bibitem{Ovalle:2017wqi}
J.~Ovalle, R.~Casadio, R.~da Rocha and A.~Sotomayor,
%``Anisotropic solutions by gravitational decoupling,''
Eur. Phys. J. C \textbf{78} (2018) 122

\bibitem{daRocha:2021aww}
R.~da Rocha,
%``Gravitational decoupling and superfluid stars,''
Eur. Phys. J. C \textbf{81} (2021) 845

\bibitem{daRocha:2021sqd}
R.~da Rocha,
%``Gravitational decoupling of generalized Horndeski hybrid stars,''
Eur. Phys. J. C \textbf{82} (2022) 34

\bibitem{Contreras19} E. Contreras, \'A.~Rinc\'on, P. Bargueno, \textit{Eur. Phys. J. C} \textbf{2019}, 79, 216.

\bibitem{Estrada19} M. Estrada, \textit{Eur. Phys. J. C} \textbf{2019}, 79, 918.

\bibitem{Gabbanelli18} L. Gabbanelli, \'A.~Rinc\'on, C. Rubio, \textit{Eur. Phys. J. C} \textbf{2018}, 78, 370.

\bibitem{Leon21} P. Leon, A. Sotomayor, \textit{Fortschr. Phys.} \textbf{2021}, 69, 1051.

\bibitem{MauryaFRT} S. K. Maurya, F. Tello-Ortiz, S. Ray, \textit{Phys. Dark Universe} \textbf{2021}, 31, 100753.

\bibitem{Maurya20} S. K. Maurya, \textit{Eur. Phys. J. C} \textbf{2020}, 80, 429.

\bibitem{Panotopoulos18} G. Panotopoulos, \'A.~Rinc\'on, \textit{Eur. Phys. J. C} \textbf{2018}, 78, 851.

\bibitem{SharifFG} M. Sharif, S. Saba, \textit{Int. J. Modern Phys. D} \textbf{2020}, 29, 2050041.

\bibitem{Zubair21a} M. Zubair, M. Amin, H. Azmat, \textit{Physica Scripta} \textbf{2021}, 96, 125008.

\bibitem{Zubair21b} M. Zubair, H. Azmat, M. Amin, \textit{Chin. J. Phys.} \textbf{2021}, 77, 898.

\bibitem{Muneer:2021lfz}
Q.~Muneer, M.~Zubair and M.~Rahseed,
%``Gravitational decoupled anisotropic spherical solutions in f(R, T) gravity by minimal geometric deformation approach,''
Phys. Scripta \textbf{96} (2021) no.12, 125015

\bibitem{Zubair:2021zqs}
M.~Zubair, M.~Amin and H.~Azmat,
%``Anisotropic charged Heintzmann solution using gravitational decoupling through extended geometric deformation approach,''
Phys. Scripta \textbf{96} (2021) no.12, 125008

\bibitem{Azmat:2021qig}
H.~Azmat and M.~Zubair,
%``An anisotropic version of Tolman VII solution in $f(R, T)$ gravity via gravitational decoupling MGD approach,''
Eur. Phys. J. Plus \textbf{136} (2021) no.1, 112

\bibitem{Abellan:2020dze}
G.~Abell\'an, \'A.~Rinc\'on, E.~Fuenmayor and E.~Contreras,
%``Anisotropic interior solution by gravitational decoupling based on a non-standard anisotropy,''
Eur. Phys. J. Plus \textbf{135} (2020) 606

\bibitem{Maurya21a} S. K. Maurya, et al., \textit{Eur. Phys. J. C} \textbf{2021}, 81, 848.

\bibitem{Maurya21b} S. K. Maurya, et al., \textit{Fortschr. Phys.} \textbf{2021}, 69, 2100099.

\bibitem{Maurya22} S.~K.~Maurya, K.~N.~Singh, M.~Govender and S.~Hansraj,
%``Gravitationally Decoupled Strange Star Model beyond the Standard Maximum Mass Limit in Einstein\textendash{}Gauss\textendash{}Bonnet Gravity,''
Astrophys. J. \textbf{925} (2022) 208

\bibitem{Casadio:2019usg}
R.~Casadio, E.~Contreras, J.~Ovalle, A.~Sotomayor and Z.~Stuchlick,
%``Isotropization and change of complexity by gravitational decoupling,''
Eur. Phys. J. C \textbf{79} (2019) 826

\bibitem{Contreras:2022vmk}
E.~Contreras, E.~Fuenmayor and G.~Abell\'an,
%``Uncharged and charged anisotropic like-Durgapal stellar models with vanishing complexity,''
Eur. Phys. J. C \textbf{82} (2022)  187

\bibitem{Carrasco-Hidalgo:2021dyg}
M.~Carrasco-Hidalgo and E.~Contreras,
%``Ultracompact stars with polynomial complexity by gravitational decoupling,''
Eur. Phys. J. C \textbf{81} (2021)  757
\bibitem{Contreras:2021xkf}
E.~Contreras and E.~Fuenmayor,
%``Gravitational cracking and complexity in the framework of gravitational decoupling,''
Phys. Rev. D \textbf{103} (2021)  124065

\bibitem{Maurya:2022cyv}
S.~K.~Maurya, A.~Errehymy, R.~Nag and M.~Daoud,
%``Role of Complexity on Self-gravitating Compact Star by Gravitational Decoupling,''
Fortsch. Phys. \textbf{70} (2022)  2200041

\bibitem{Maurya:2022pef}
S.~K.~Maurya, M.~Govender, S.~Kaur and R.~Nag,
%``Isotropization of embedding Class I spacetime and anisotropic system generated by complexity factor in the framework of gravitational decoupling,''
Eur. Phys. J. C \textbf{82} (2022)  100

\bibitem{Maurya:2021yhc}
S.~K.~Maurya and R.~Nag,
%``Role of gravitational decoupling on isotropization and complexity of self-gravitating system under complete geometric deformation approach,''
Eur. Phys. J. C \textbf{82} (2022)  48

\bibitem{Abbott19} B. P. Abbott, et al., \textit{Phys. Rev. X} \textbf{2019}, (LIGO Scientific Collaboration \& Virgo Collaboration), 9, 031040.

\bibitem{Abbott20b} R. Abbott, et al., \textit{Phys. Rev. D} \textbf{2020}, 102, 043015.

\bibitem{Abbott20a} B. P. Abbott, et al., \textit{The Astro. J. Lett.} \textbf{2020}, 892, L3.

\bibitem{Dimakis21} N. Dimakis, A. Paliathanasis, T. Christodoulakis, \textit{Class. Quantum Grav.} \textbf{2021}, 38, 225003.

\bibitem{Chodos:1974} A. Chodos, R. L. Jaffe, K. Johnson, C. B. Thorn, V. F. Weisskopf, \textit{Phys. Rev. D} \textbf{1974}, 9, 3471.

\bibitem{Farhi:1984} E. Farhi, R. L. Jaffe, \textit{Phys. Rev. D} \textbf{1984}, 30, 2379.

\bibitem{cap} A. Bauswein, et al., \textit{The Astro. J. Lett.} \textbf{2017}, 850, L34.

\bibitem{bau} C. D. Capano, et al., \textit{Nat. Astron.} \textbf{2020}, 4, 625.

\bibitem{star1} P. Demorest, T. Pennucci, S. Ransom, et al., \textit{Nature} \textbf{2010}, 467, 1081.

\bibitem{star2}  P. C. C. Freire, C. G. Bassa, N. Wex, et al., \textit{Mon. Notices Royal Astron. Soc.} \textbf{2011}, 412, 2763.

\bibitem{star3} M. L. Rawls, J. A. Orosz, J. E. McClintock, M. A. P. Torres, C. D. Bailyn, M. M. Buxton, \textit{The Astro. J} \textbf{2011}, 730, 25.

\bibitem{wenbin} W. Lu, P. Beniamini, C. Bonnerot, \textit{Mon. Notices Royal Astron. Soc.} \textbf{2021}, 500, 1817.

\end{thebibliography}
\end{document}